\documentclass{article}
\usepackage{amsmath,amssymb}
\usepackage{graphicx}
\usepackage{hyperref}
\usepackage[right]{lineno}

\usepackage[aboveskip=1pt,labelfont=bf,labelsep=period,justification=raggedright,singlelinecheck=off]{caption}

\makeatletter
\renewcommand{\@biblabel}[1]{\quad#1.}
\makeatother

\date{}

\providecommand{\zfff}{\textit{zf4f}}
\providecommand{\gill}{Gill et al.}
\providecommand{\pillow}{Pillow et al.}

\begin{document}
\vspace*{0.35in}

\begin{flushleft}
{\Large
\textbf\newline{Detailed temporal structure of communication networks in groups of songbirds}
}
\newline
\\
Dan Stowell\textsuperscript{1*},
Lisa Gill\textsuperscript{2},
David Clayton\textsuperscript{3}
\\
\bigskip
\bf{1} Centre for Digital Music, Queen Mary University of London, UK
\\
\bf{2} Max Planck Institute for Ornithology, Seewiesen, Germany
\\
\bf{3} School of Biological and Chemical Sciences, Queen Mary University of London, UK
\\
\bigskip

* dan.stowell@qmul.ac.uk

\end{flushleft}

\section*{Abstract}
Animals in groups often exchange calls, in patterns whose temporal structure
may be influenced by contextual factors such as physical location and the social network structure of the group.
We introduce a model-based analysis for temporal patterns of animal call timing,
originally developed for networks of firing neurons.
This has advantages over cross-correlation analysis in that it can correctly handle common-cause confounds
and provides a generative model of call patterns with explicit parameters for the influences between individuals.
It also has advantages over standard Markovian analysis in that it incorporates detailed temporal interactions which affect timing as well as sequencing of calls.
Further, a fitted model can be used to generate novel synthetic call sequences.
We apply the method to calls recorded from groups of domesticated zebra finch (\textit{Taenopyggia guttata}) individuals.
We find that the communication network in these groups has stable structure that persists from one day to the next,
and that ``kernels'' reflecting the temporal range of influence have a characteristic structure for a calling individual's effect on itself, its partner, and on others in the group.
We further find characteristic patterns of influences by call type as well as by individual.

\vspace{2cm}

\textbf{Keywords:}
animal communication;
Poisson process;
point process;
\newline
linear-nonlinear Poisson;
communication network;
social network analysis.


\section*{Introduction}
\label{sec:intro}

Many animals exhibit group calling behaviour.
Patterns of calling are observable phenomena which reflect individual state
and are dependent on behavioural context \cite{Elie:2015,Gill:2015}.
Understanding the dynamics of vocalisation patterns within groups is an important growing topic in animal behaviour \cite{Greenfield:2006,Kershenbaum:2014b,Gill:2015,Perez:2015}.
However, analysing the structure of the communication network in a group of animals presents a challenge which goes beyond that of analysing calls of isolated individuals or pairs,
because multiple influences converge on an individual in parallel,
making it harder to infer causal connections.

In this work we introduce a model-based method for inferring the temporal and network structure of interactions between calling individuals
from the timing of call events.
The paradigm was originally developed in computational neurology for analysis of spiking neural networks \cite{Paninski:2004,Pillow:2008}.
We adapted the method for the case of animal calls
and applied it to data from groups of domesticated zebra finch (\textit{Taenopyggia guttata}),
a communal songbird that is the subject of much current research \cite{Zann:1996,Warren:2010}.
With this approach we were able to represent zebra finch communication networks in a compact model whose attributes reflect fine details of timing and influence strengths between individuals in a group,
yielding a new data-driven perspective that complements other approaches based on acoustics, neurology or ethology,
and provides a useful visualisation tool.

Before describing our study and analysis,
we first wish to set our analytical approach in context by discussing methods for modelling animal vocalisation sequences,
in particular their applicability to vocalisations in groups.

\subsection*{Modelling the Processes that Generate Animal Vocalisations}

Researchers analyse animal calling patterns in order to understand the processes that generated them,
whether their focus is on intra-individual or inter-individual mechanisms.
A general paradigm with strong mathematical support is to choose a family of probabilistic generative models that might generate the phenomena of interest, and then to use model selection and/or parameter fitting to decide which model from that family best matches the data.

A good example of this is Markov modelling. A Markov model generates the next symbol in a sequence stochastically but with limited memory: conditional on the most recent $k$ symbols, a $k$th-order Markov model chooses the next symbol independently of all prior history. Markov modelling has been applied widely to animal communications \cite{Kershenbaum:2014b};
once vocalisations have been reduced to symbol sequences, data fitting can determine the transition probabilities between symbols, as well as the model order i.e. the length $k$ of the ``memory'' \cite{Briefer:2010}.
A Markov model is usually an extreme simplification of the presumed underlying biological process, and neglects important aspects such as call timing.
Because of this, a Markov model is unable to model some notable aspects of vocalisation such as ``bursty'' call patterns.
However it is a broadly useful tool.
Extensions of this approach augment the model to include unobserved state (the \textit{hidden Markov model} long used for speech \cite{Rabiner:1993}, \cite[Chapter 17]{Murphy:2012}), the structured repetition of symbols (the \textit{semi-Markov model} \cite{Kershenbaum:2014,Stowell:2014c}) or time gaps between events (the \textit{Markov renewal process} \cite{Ball:2005,Stowell:2013,Lasko:2014}).
For our present purposes, an important consideration is that Markovian models do not adapt readily to the case where multiple influences converge on an individual in parallel.
They imply that each individual ``remembers'' only the most recent $k$ calls, irrespective of whether they come from socially significant others (e.g. a breeding partner) or from socially insignificant individuals.

\begin{figure*}[tp]
	\caption{A synthetic example of indirect causation. A sequence was generated using a simplified A$\to$B$\to$C causal model, and then analysed using standard methods and our proposed method. The generation procedure was deliberately designed as not a perfect match to any of the analysis models, so as not to privilege any of them. We show an example timeline of events, along with the cross-correlation plots, and then the influence strengths recovered using each method summarised as a social network diagram. Cross-correlation analysis recovers most of the influences/independences but tends to recover false-positive connections for the indirect link A$\to$C (see the lower-left panel of the cross-correlation plots). A simple Markov model recovers influences without timing information, and in this case also adds a connection from C back to A to take the place of the baseline event calling rate of A. Our proposed method recovers a good match for the network structure as well as timing information. It adds self-inhibitory feedback on B and C to account for the fact that in this test case a call by A leads to no more than one call by B (and likewise for B$\to$C). Note that the values recovered by each method are different in kind, and have been rescaled separately for each of the network plots. For further details of this synthetic example see the SI.}
	\centering
	\includegraphics[width=0.99\linewidth]{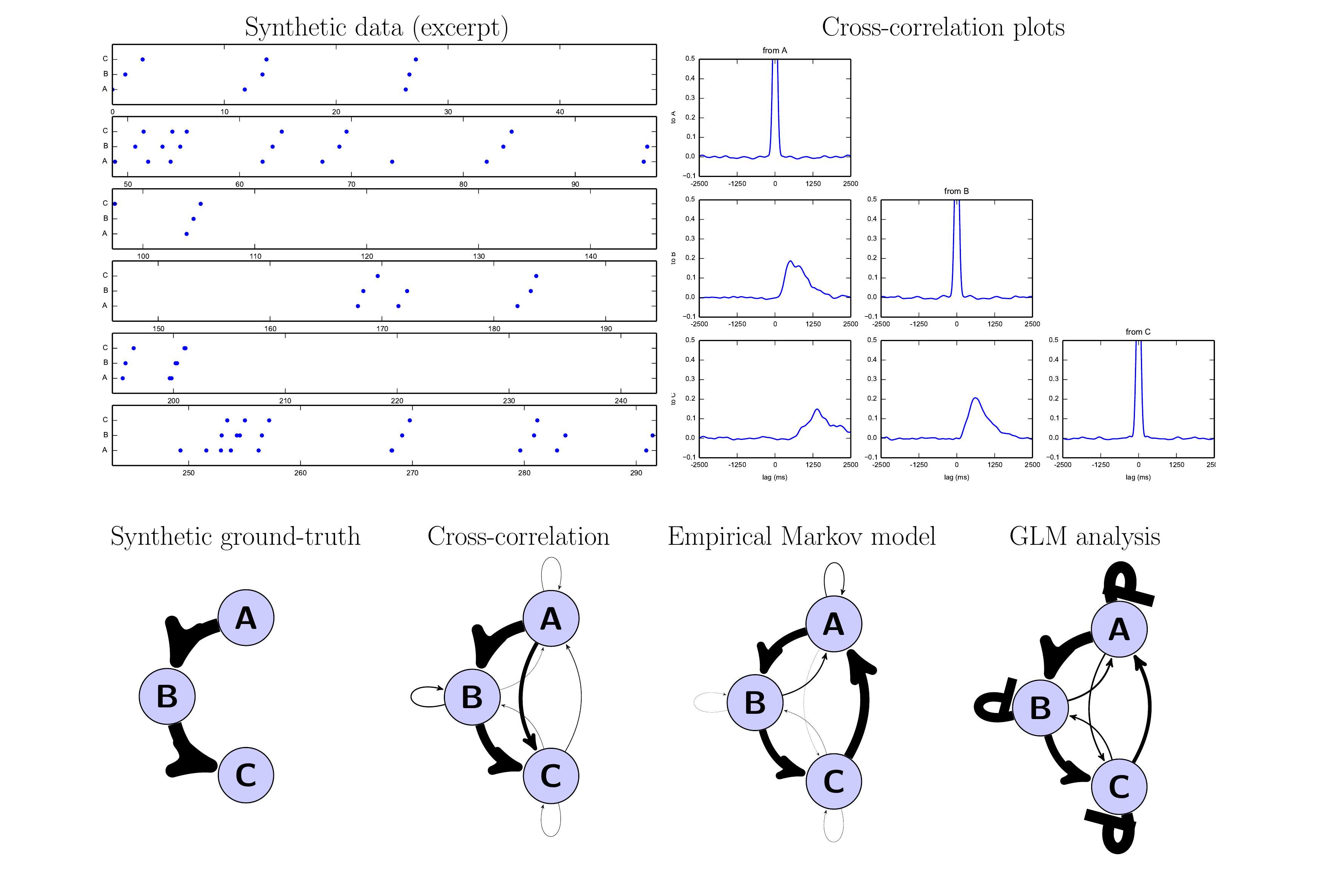}
	\label{fig:abcexample}
\end{figure*}

Cross-correlation analysis can be used to analyse relative timing, but is not derived from a generative model: it is descriptive rather than inferential.
A particular set of cross-correlation statistics may be compatible with multiple hypotheses about the underlying process (Markovian or otherwise).
Cross-correlation is an appealing alternative to standard Markov modelling because it gives some characterisation of the time gaps between events, not just the event sequences.
Studies based on cross-correlation typically probe for significant patterns but do not attempt to give a formal model that could have generated those patterns \cite{TerMaat:2014,Gill:2015}.
As one example of potential issues with cross-correlation, a causal network with a chain structure such as A$\to$B$\to$C may well create indirect cross-correlation phenomena from A$\to$C, even where there is no direct causal link (Fig. \ref{fig:abcexample}),
which would result in a clear instance of the maxim ``correlation is not causation''.

\begin{figure}[t]
	\caption{Schematic illustration of processes generating events: (a) homogeneous Poisson process; (b) inhomogeneous Poisson process; (c) inhomogeneous Poisson process in which all changes in rate are due to the external influence of stimulus events. In each panel, the rate parameter for the process ($\lambda$) is shown as a filled curve, continuous in time, and an example sequence of events sampled from the process is shown as a sequence of spikes.
}
	\centering
	\includegraphics[width=0.6\linewidth]{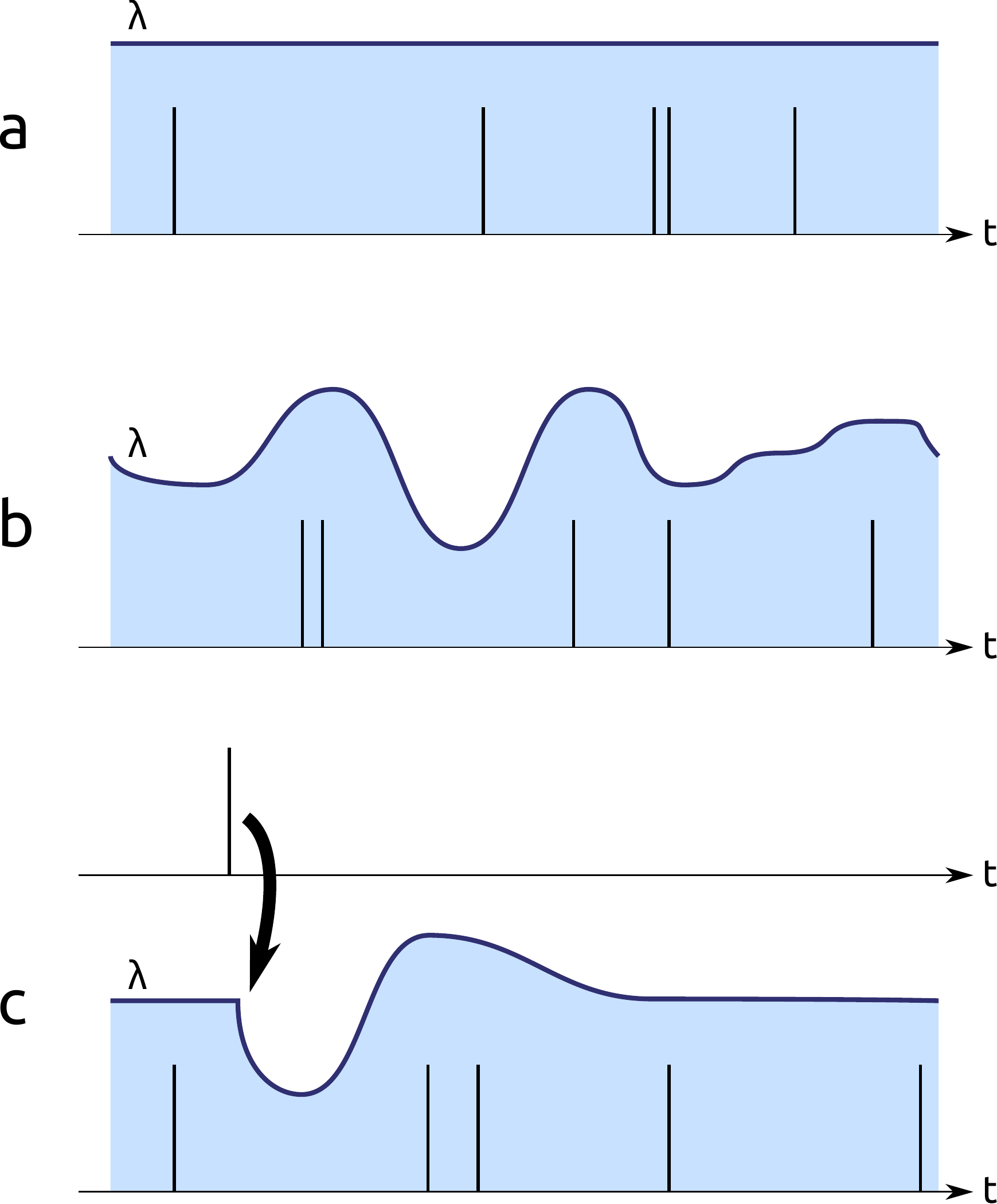}
	\label{fig:spikingprocesses}
\end{figure}

A probabilistic model that is directly applicable to events on a continuous timeline is the \textit{Poisson process}
\cite{Daley:1988,Dayan:2001}.
At its simplest, the (homogeneous) Poisson process outputs events stochastically but at a constant rate, meaning that the event times are random but there is a constant expected number of events per unit time (Fig. \ref{fig:spikingprocesses}a).
Most Poisson processes of interest are inhomogeneous, having a rate that can change over time (Fig. \ref{fig:spikingprocesses}b).
This Poisson process model can represent a single stream of events, but in order to capture interactions between individuals or between call types, we need to augment it with coupling such that calls from one individual can modulate the rate of calling of another (Fig. \ref{fig:spikingprocesses}c).
This will be achieved through influence \textit{kernels} described in the SI and illustrated later,
whose effect is that a call from one individual has a modulating `wave' of influence on the rates of others.
Such a model can be fit to data by maximum likelihood and these models reflect both the typical time gaps between events and the typical sequencing of one event after another.
These coupled processes are not necessarily Markovian, and the modelling focus is slightly different: instead of the turn-by-turn sequencing which underlies Markovian models,
the emphasis here is on separate processes each generating calls, happening in parallel, and these processes can mutually influence one another.

In the present study we wished to use exactly these point process models, with interaction kernels, to elucidate the temporal structure of networks of calling birds.
We have multiple motivations for doing this.
The first is that the information from the fitted model may yield information similar to that which has previously been derived through cross-correlation \cite{Perez:2015,Gill:2015}, such as the strength of pairwise influences, but with more robustness to common-cause confounds and other issues discussed above.
A second is that the fitted model may yield more finely nuanced information, and information tied to an explicit model:
for example, one individual may have a suppressive effect on its neighbour at some timescales, and an excitatory effect at other timescales.
Another is that a fitted model can be used to make further inferences about datasets,
for example to predict whether communicating partners are paired or unpaired.
Another is that since the model we fit is generative, it can be used to generate new synthetic sequences having the same network characteristics,
which could be used for stimuli in future studies.

To this end, we conducted one study with a group of female zebra finches in a standardised context,
and one reanalysis of existing data from a mixed-sex group of zebra finches in a different and varying context.

\section*{Materials and Methods}
\label{sec:methods}

\subsection*{Ethical Note}

Animal housing and welfare were in compliance with the European directives for the protection of animals used for scientific purposes (2010/63/EU).
Our zebra finch audio recordings did not involve any intervention that would be a regulated procedure under the UK Animals (Scientific Procedures) Act 1986. The Named Animal Care and Welfare Officer for Queen Mary University of London (QMUL) oversaw the housing and routine care. 

In this report we also reanalyse data from \gill \cite{Gill:2015}. Those data were collected under protocols approved by the Government of Upper Bavaria, and conditions were likewise compliant with 2010/63/EU.

\subsection*{Data Collection}

Four adult female zebra finches (at least 90 days old, with wild-type plumage) were selected from an aviary of the QMUL animal facility. The four birds were housed together in a flight cage with free access to food and water, in a room separate from the main aviary. The group of four birds were housed together for more than two weeks before the recording sessions. The birds were kept on a 12/12h light/dark cycle (7:00--19:00), and the room temperature was 20--21$^{\circ}$C.

To perform the recordings, each bird was transferred to an individual cage (of size approx.\ 40 x 35 x 45 cm) with free access to food and water, and remained in visual and auditory contact with the other birds (at a distance of about 2 m).
Birds were kept in the individual cages for just over one hour per day (approx 8:00--9:00) for recordings before returning to the group cage.

The solo cages were arranged in a square pattern so that all birds were approximately equidistant. Since we intended to investigate calling patterns as a function of bird identity, we avoided the potential confound of physical location by placing the birds in cages in different orders each of the three days, choosing the ordering by taking three rows from a four-by-four Latin square.

Audio was recorded during these one-hour sessions with four focal microphones (AKG C451B), one directed at each cage.
All audio signals were recorded together onto a Zoom H6n multitrack sound recorder to ensure that the recordings were temporally synchronised.
Recordings were made at 96 kHz sample rate and 16 bit depth.
The first day of this sound recording protocol was used as a test run and for acclimation, and data were not analysed.
The second and third days were taken forward for annotation.

We used a cross-validated semi-automatic process to label the audio events in the recordings. In a first pass, we applied automatic event detection to locate the beginning and end of events, using energy-based detection applied to spectrograms after performing median-filtering. Spectrograms were trimmed to the frequency region of interest (0.5--20 kHz).
The four channels of the recording inevitably contained large amounts of ``crosstalk'' as the protocol was designed so that the birds could clearly hear each other. Hence we used the median spectrogram across all channels as a background against which to judge the signal energy.
Regions of high energy exceeding a minimum duration (8 ms; 2 spectrogram frames or more) were taken forward as candidates to the second stage.

The second stage of processing was manual refining of the detections. Each 4-channel spectrogram was divided into one-second chunks (with an overlap of 50\%) and the proposed annotation superimposed.
Annotators were then shown the one-second chunks in a random order, so that any variation in attention of the human annotators would not systematically vary across the recording duration.
For each channel, annotators could listen to the audio and view spectrograms, confirm or reject the detected events, and label them as calls or other noises (such as wing flaps or cage sounds).
They could also label errors such as the merging or splitting of events, or misaligned event boundaries.
Two separate annotators manually processed all of the candidate annotations.
We used two paid annotators who were both PhD students with critical listening skills (audio engineering backgrounds).
These annotators were trained by the first author, using excerpts from the pilot session recordings as examples.
Consensus decisions were accepted, and deviations from consensus were resolved by the first author.
Finally, the first author listened through to the full recordings with the annotations superimposed,
as a check for any remaining anomalies.

For the present study we then extracted the calling times for those events labelled as zebra finch calls.
In this standardised recording environment our female birds did not use a wide variety of call types (cf.\ \cite{Zann:1996}), predominantly the `Tet'/`Stack' type with a very small number of `Distance' calls (note that female zebra finches do not sing).
Therefore in order to provide a clear analysis we did not split calls into different categories.
This dataset of call times we refer to as \zfff.%
\footnote{Data available from: \url{https://dx.doi.org/10.6084/m9.figshare.1613791}}
In this dataset there were around 2800 calls in each hour-long session (around 12 calls per bird per minute).

\subsection*{Analyses}

We analysed the call timing information using the GLM point-process model of \cite{Pillow:2008},
with specific configuration described in the SI.
Our main unit of analysis was each 60-minute recording session as a whole,
although we also analysed each 15-minute segment separately,
to investigate whether there was continuity or change of communication patterns throughout a session.
We determined from preliminary tests that 15 minutes was the smallest region we could use to have enough calls for a stable analysis.

We used penalised maximum likelihood (maximum \textit{a posteriori}) optimisation to fit the model parameters to each dataset. 
We fitted two models to each of our 60-minute sessions---one combining influences in additive fashion, one in multiplicative fashion---and used an odds-ratio test to select the most appropriate model.
In all cases the additive model was favoured.
More details on the modelling are given in the Supplementary Information.
Our source code to perform analyses and generate figures is available online.%
\footnote{Software available from: \url{https://bitbucket.org/danstowell/callnets-glm}}

The GLM model, after fitting to a dataset, yields a continuous curve (a ``kernel'') for each directed pairwise influence.
For four birds this gives 16 kernels, four of which represent self-self influence and twelve of which represent self-other influence.
In order to make quantitative and visual comparisons, we summarised kernels in two main ways.
We aggregated self-self and self-other kernels separately to look for general tendencies that might emerge independent of bird identity.
Separately we used the magnitude of the strongest peak (positive or negative) as summary statistics to plot communications networks and to test for consistency within/between sessions.

To test for consistency between sessions, we took the peak strengths from the influence kernels, and measured the Pearson correlation from one day to the next.
Since self-self and self-other kernels were different in kind (having positive and negative peaks respectively),
we analysed them separately, giving $N=12$ for self-other and $N=4$ for self-self.
To investigate whether any correlation was due to individual identity, to physical location of the cage, or to chance,
we measured the correlation using four different ways of matching one day up with the next day.
These four matchings were the combinations given by a Latin square:
one matching compared the same individual across days,
one matching compared the same physical location (and microphone) across days,
while the remaining two were null matchings with no meaningful interpretation.

To test for consistency within sessions, we took the self-other peak strengths for each 15-minute segment,
and measured via Pearson correlation how strongly a segment could predict the immediate next segment.
For each one-hour session this gives three sequential pairs of segments.

All of the above analysis was applied to our own \zfff dataset.
We also applied the analysis to a subset of the call data from \cite{Gill:2015}
in which inter-individual call timings had been studied in a more complex group setting.
These groups were made up of freely behaving zebra finch males (4) and females (4) whose vocalisations were individually recorded via backpack microphones, over a period of three weeks.
During this time, all birds were able to interact physically, and to engage in various activities,
while pair-bonding, nesting and breeding progressed.
We used our GLM method as an alternative to the cross-correlation analysis of \gill,
to investigate group calling behaviour on a finer timescale, including an investigation of self-self calling patterns.
As all individuals in the analysed data formed pair-bonds,
we not only distinguished between self-self and self-other interactions as in \zfff, but
we also separated out the self-partner interactions to investigate any consistent patterns emerging specifically within pairs.
We were provided with the data for Trial II, days 1, 7, 11, 18%
---the same days as displayed in Fig. 5A of \cite{Gill:2015}---%
which span the different breeding stages of that group.%
\footnote{Data available from: \url{https://figshare.com/s/73cbdf96ab156a0f0a69}}
Each session was just under four hours long.

In the data of \gill \cite{Gill:2015}, calls were categorised into types.
We applied our GLM method to all calls together (as with our own data),
but since this dataset contained a larger number of calls across various types,
we also explored splitting the data according to call type so that each pairwise interaction could have a different kernel for each call type.
Note that \gill\ use five call categories, which implies that for each directed pair we must fit 25 different kernels rather than just one, and so in practice we found that the full analysis by call-type was less stable due to data sparsity issues.
Hence the recovered per-type kernels showed larger variances and the results are in some cases less clear-cut than for the broader aggregate models.

\section*{Results}
\label{sec:results}

\subsection*{All-Female Group (\zfff)}

We first show results from our recordings with four female zebra finches in a standardised context.
The model-selection test consistently selected the model in which influences were combined by addition (rather than multiplication).
This is in contrast to \pillow \cite{Pillow:2008} applying the same method to an analysis of spiking neurons, which selected the multiplicative model.

\begin{figure*}[pt]
	\caption{Aggregate view of the influence kernels recovered from our two study sessions with four female zebra finches. %
%
\newline
\textbf{HOW TO READ THESE PLOTS:} These plots summarise the kernels across the entire communication network, grouping the kernels according to whether they represent a self-self (red lines) or self-other (blue) connection.
For each directed pair of birds we inferred a single kernel curve; these plots show the median curve, and the 5-to-95-percentile range, across all possible pairs of individuals in the group.
Hence the filled regions largely indicate the extent of variation among the network connections.
The time on the x-axis can be thought of as similar to the ``lag'' in cross-correlation.
The y-axis can be thought of as the ``excess calling rate'' caused by a stimulus (although this interpretation is complicated a little by the nonlinearity; see SI for detail).
Imagine that a bird emits a call at time zero. The plot then shows the effect of that call over the next few seconds,
increasing and/or decreasing every bird's tendency to call.
Unlike a Markov model, the call at time zero is not considered to lead to a single call that happens as a consequence of it: another bird might call once, twice or more during the period in which it is strongly stimulated by the call at time zero. (In practice, the strong inhibition we see---the strong negative peak for self-self interactions---often suppresses multiple responding.)
A flat kernel with a value of zero would correspond to statistical independence, indicating that one bird had no effect on the calling rate of the listener.
The influences from multiple individuals are added together by the listener
before being passed through a nonlinearity;
the main effect of the nonlinearity (for interpretive purposes) is a soft-thresholding to prevent the rate going below zero.
For self-self kernels, the lag includes the lag due to the duration of the call itself (median duration 0.1 s).
}
	\centering
	\includegraphics[width=0.99\linewidth,clip,trim=0mm 0mm 0mm 0mm]{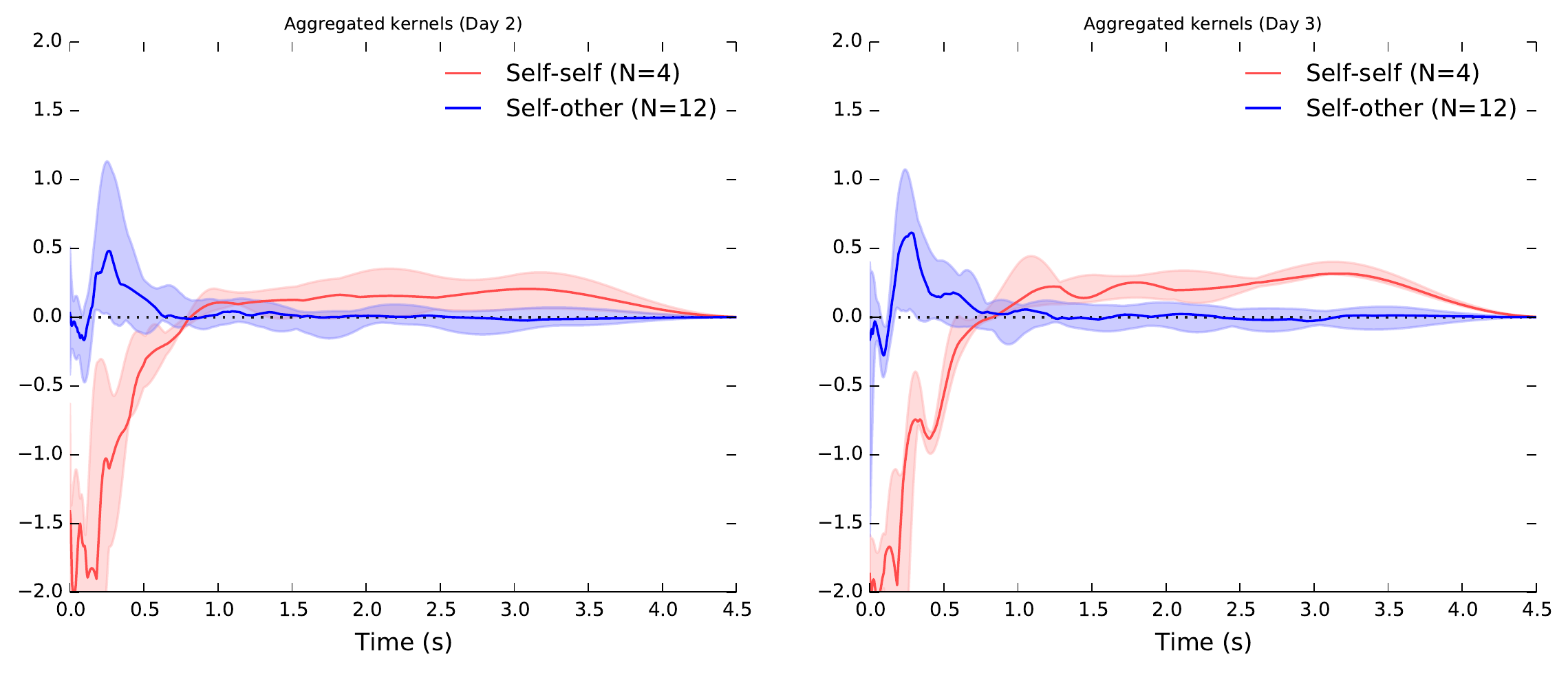}
	\label{fig:plot_kernels_ssso}
\end{figure*}

\begin{figure}[t]
	\caption{Peak influence strengths, plotted as a social network. Standard arrows indicate postive (excitatory) peaks, flat-headed arrows (in this case only seen for the self-self arrows looping back) indicate negative (suppressive) peaks.}
	\centering
	\includegraphics[width=0.8\linewidth,clip,trim=105mm 155mm 105mm 2mm]{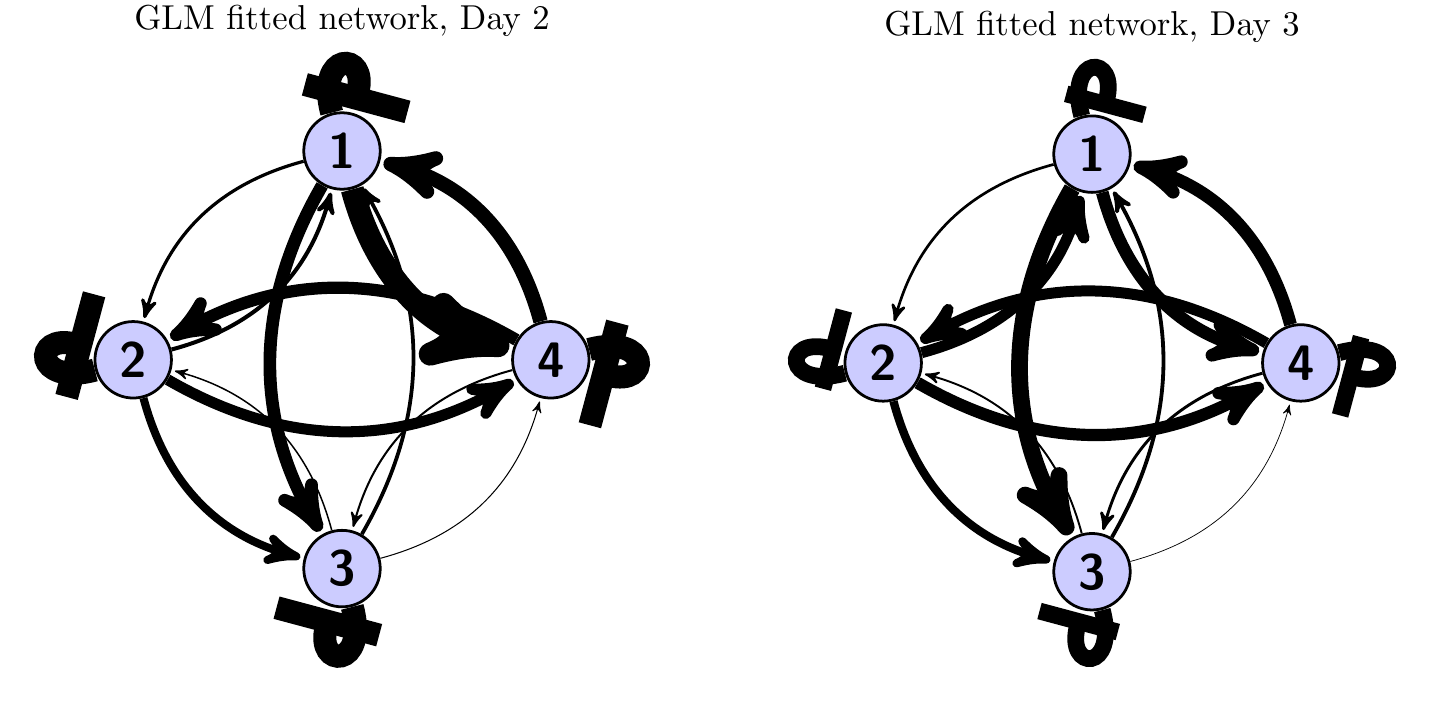}
	\label{fig:netplot_callnets}
\end{figure}

We found that influence kernels exhibited consistent temporal characteristics and broadly consistent magnitudes, and that these differed very strongly between self-self and self-other interactions (Fig. \ref{fig:plot_kernels_ssso}).
The confidence intervals, despite aggregating over individuals, are relatively narrow
with little overlap between self-self and self-other.
Individuals exhibited a pattern of strong self-suppression immediately after calling and for around the next 0.8 seconds,
followed by a slight positive effect thereafter.
In contrast, self-other interactions showed a consistent positive peak at around 0.25 seconds, before decaying to around zero at 0.7 seconds,
indicating a consistent characteristic timescale for calls that occur in response to the calls of others.
In this group of females, although the network influences showed consistency there was no evidence for strong structure of the network such as a hierarchy (Fig. \ref{fig:netplot_callnets}).

The self-other peak strengths from one day to the next were strongly predicted by individual identity ($p < 0.001$),
and not by physical location or by the null combinations (Table \ref{tbl:plot_analyseadj_between}).
Physical location yielded a slightly stronger correlation than the null combinations,
but not at a significant level ($p = 0.18$).
Thus, we attribute the variation in inter-individual peak influence strengths (Fig. \ref{fig:netplot_callnets}) to individual identity.
The same was true of self-self peak strengths, predictable by individual identity ($p < 0.01$) but not by the other permutations ($p > 0.05$).

\begin{table}[t]
\centering
\caption{Predictability (Pearson correlation) of peak strengths, from one day to the next, measured under four different permutations for aligning the two days. The four permutations correspond to four rows of a Latin square, one of which matched individuals across days, another which matched physical cage locations across days, and two null permutations having no meaningful interpretation.}
\label{tbl:plot_analyseadj_between}
\begin{tabular}{lllll}
Data permutation & Self-other (N=12) & Self-self (N=4) \\
\hline
Individual & \textbf{0.85 ***} & \textbf{0.99 *} \\
Location & 0.41 & 0.88 \\
Null 1 & -0.12 & -0.92 \\
Null 2 & 0.14 & -0.95
\end{tabular}
\end{table}

\begin{figure}[t]
	\caption{Inter-individual peak strengths, analysed in subsequent 15-minute time windows. Solid lines connect peak strengths for each directed pair across directly adjacent 15-minute periods, and dotted lines represent the same connection but with a gap from one day to the next.}
	\centering
	\includegraphics[width=0.89\linewidth,clip,trim=12mm 7mm 14mm 12mm]{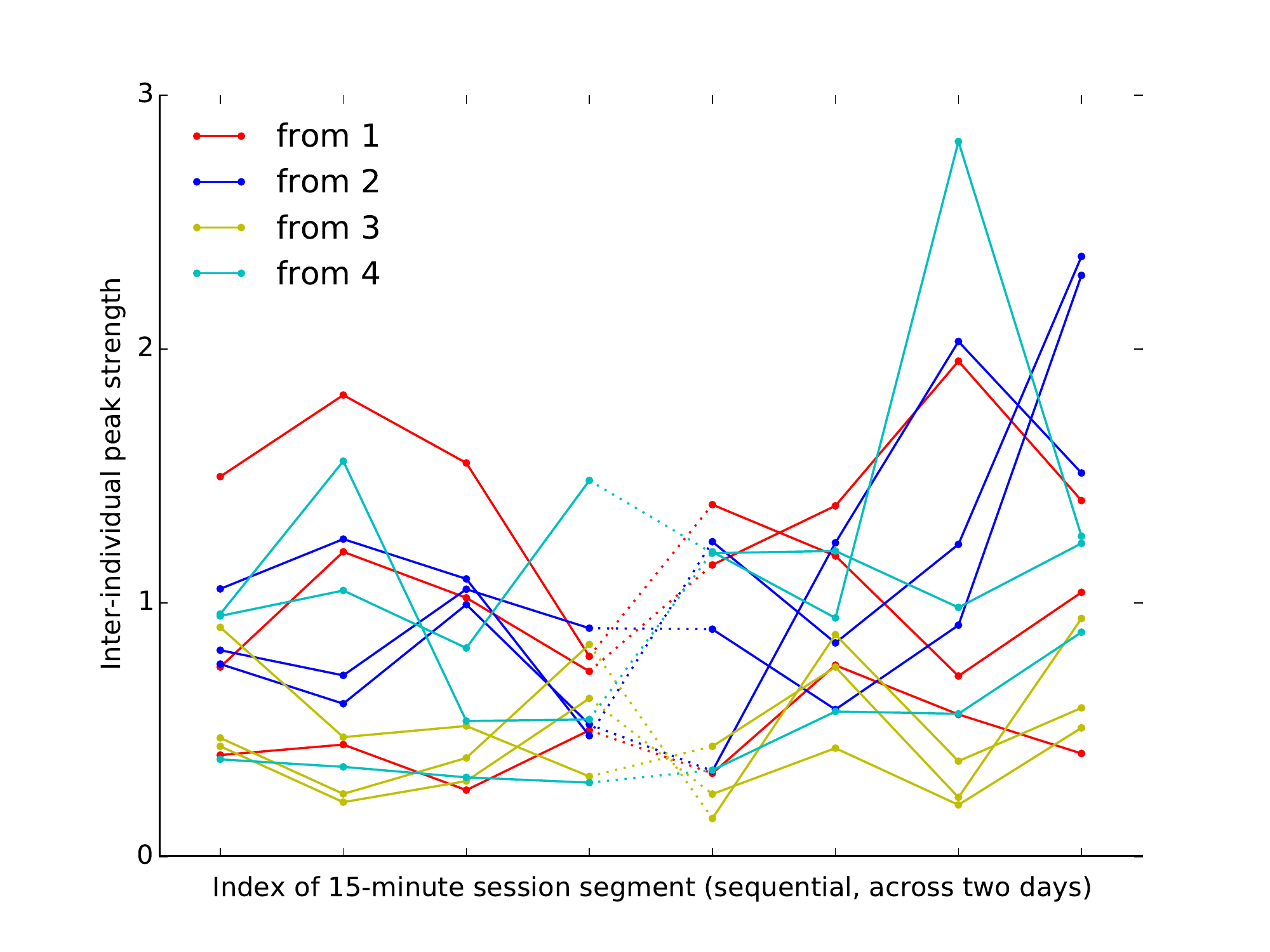}
	\label{fig:plot_analyseadj_15seq}
\end{figure}

When analysing the sessions in 15-minute segments, we found consistency but also variation in the self-other peak strengths: values did not remain constant but often were smoothly-varying with characteristic magnitudes (Fig. \ref{fig:plot_analyseadj_15seq}).
The strengths in each 15-minute segment were predictable from the immediate preceding segment (Pearson correlation 0.51, $p < 10^{-5}$; Fig. \ref{fig:plot_analyseadj_15seq}),
confirming that the between-day consistency can be observed on the finer scale of 15 minute segments despite the observable variation.
On this timescale, we were not able to verify that the self-self peaks were as consistent ($p > 0.05$).

\subsection*{Reanalysis of Gill \textit{et al.}}

\begin{figure}[pt]
	\caption{Aggregate kernels as in Fig. \ref{fig:plot_kernels_ssso} but for the dataset of \gill. %
 Note that the ``self-partner'' category was labelled retrospectively, according to the pair bonds that eventually formed. The pairings had typically not yet formed on the first day.%
}
	\centering
	\includegraphics[width=0.6\linewidth,clip,trim=0mm 0mm 0mm 0mm]{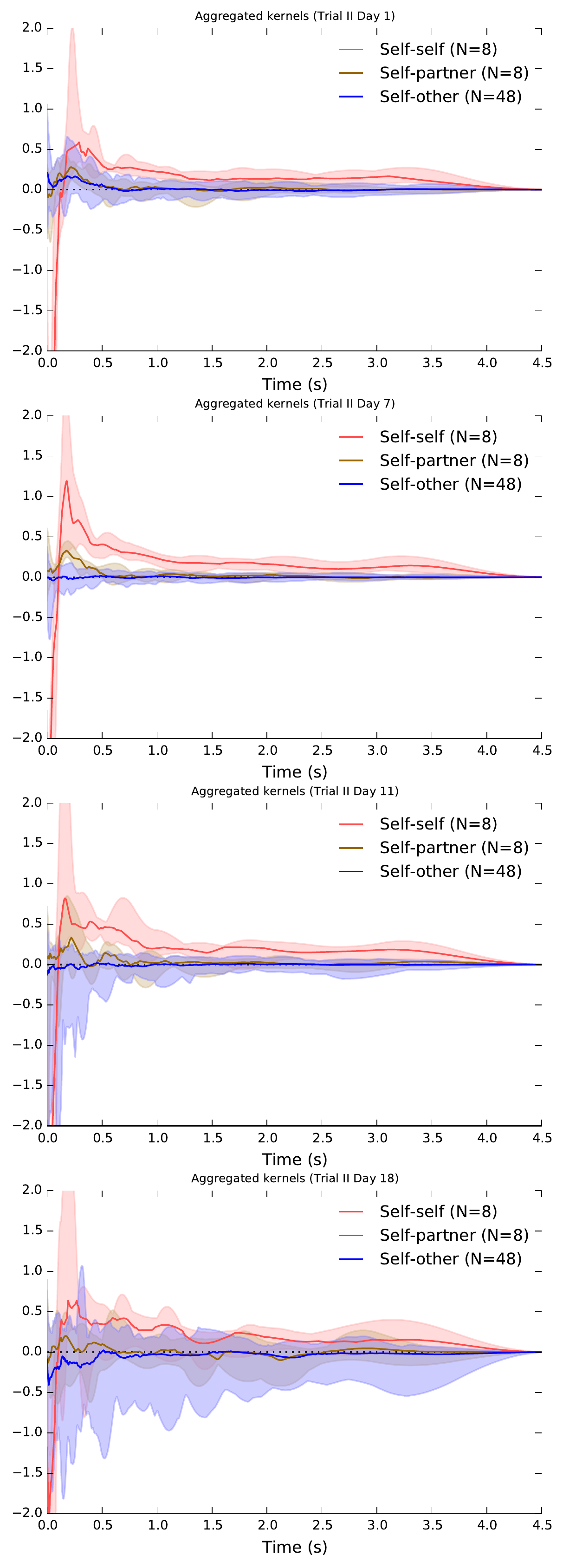}
	\label{fig:plot_kernels_ssso_gill_ind}
\end{figure}

The groups studied in \gill were in a very different environment---mixed-sex and larger groups, with the ability to physically interact, to undertake nesting and breeding.
This is reflected in some notable differences in the typical influence kernels compared against those from the \zfff data (Fig. \ref{fig:plot_kernels_ssso_gill_ind}).

Again the different kernel types show continuity over multiple sessions.
Here, however, we observed specific developments in the communication network as the pairs progressed through different stages of bonding and breeding.
On the first day, when pairings were yet to stabilise, there was little difference between self-other and self-partner influence (note that the ``self-partner'' category was labelled retrospectively, so in the early days it indicates \textit{eventual} partners).
As partnerships formed and developed they took on specific within-pair communication characteristics:
by day 7, when many of the birds were involved in nest building, communication showed a specific self-partner peak with a timescale around 0.2 s, while the typical influence of non-partner birds (self-other) had reduced down close to independence.
In other words, group communication was dominated by within-pair patterns.
In the later days this structured communication subsided somewhat,
although self-partner influences continued to be stronger than self-other influences.
In the later days, the extra-pair self-other influences showed on average a short-term suppressive effect.

Notably, the self-self influence kernels recovered from these data were rather different from those in the \zfff recordings.
There was again a strong immediate self-inhibition effect,
but in the present case this was followed by a self-excitation at around 0.2 s
which was not observed in the \zfff data.
The implication of short-term self-excitation is that calls are being emitted in bursts or sequences.
The median self-self influences showed bumpy multimodal curves which suggested that there might be further structure in the patterns of typical gaps in the sequences,
or that the aggregate plots were merging together different kernels which each had differing timescales.
We inspected the detail of individual kernel plots and found that the latter was not the case:
there were no observable individual differences in overall self-excitation timescales.

\begin{figure}[pt]
	\caption{Aggregate kernels for the dataset of \gill but showing only the within-pair interactions (self-self and self-partner) and further breaking them down by sex.}
	\centering
	\includegraphics[width=0.6\linewidth,clip,trim=0mm 0mm 0mm 0mm]{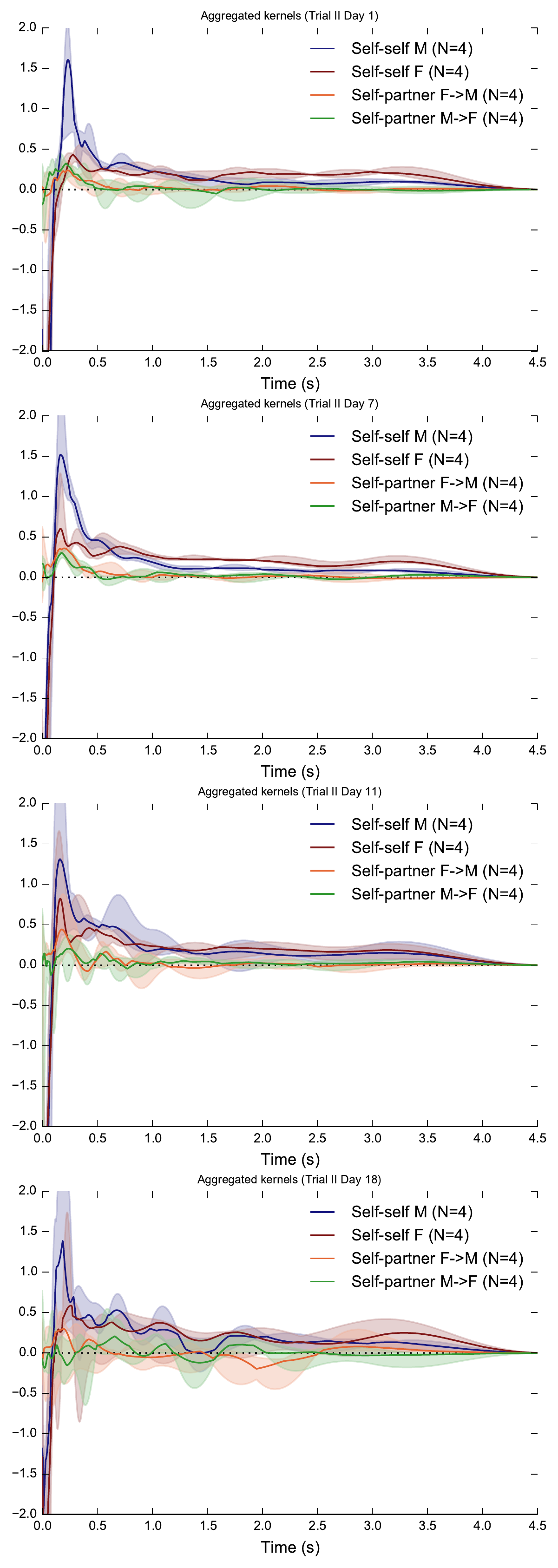}
	\label{fig:plot_kernels_ssso_gill_ind_sexwise}
\end{figure}

When inspecting the differences between males and females (Fig. \ref{fig:plot_kernels_ssso_gill_ind_sexwise}),
we found this short-term peak in self-excitation was seen more strongly in males.
The pattern became clearer when inspecting the kernels derived after separating the calls into behavioural call types.
It was observed to lie predominantly in cackle calls, specifically in an individual following a cackle with another cackle (Fig. \ref{fig:plot_kernels_ssso_gill_ind_4type_cackle}).
The kernel plots broken down by call type exhibited more variance than the main plots due to data sparsity,
but nevertheless only the cackle$\to$cackle self-self influence showed this strong rapid peak (at around 0.15 s),
and this was consistent across the different days analysed.

Cackles explained one component of the multi-modal self-self kernels.
However we could not conclude that the overall self-self kernel was explained as merely a sum of unimodal influences varying by call type, as the broken-down kernel plots did not generally resolve to simpler structure.

\begin{figure*}[pt]
	\caption{Aggregate kernels specifically for the ``Tets'' and ``Stacks'' of \gill, and their interactions, on day 7.}
	\centering
	\includegraphics[width=0.99\linewidth,clip,trim=110mm 170mm 220mm 85mm]{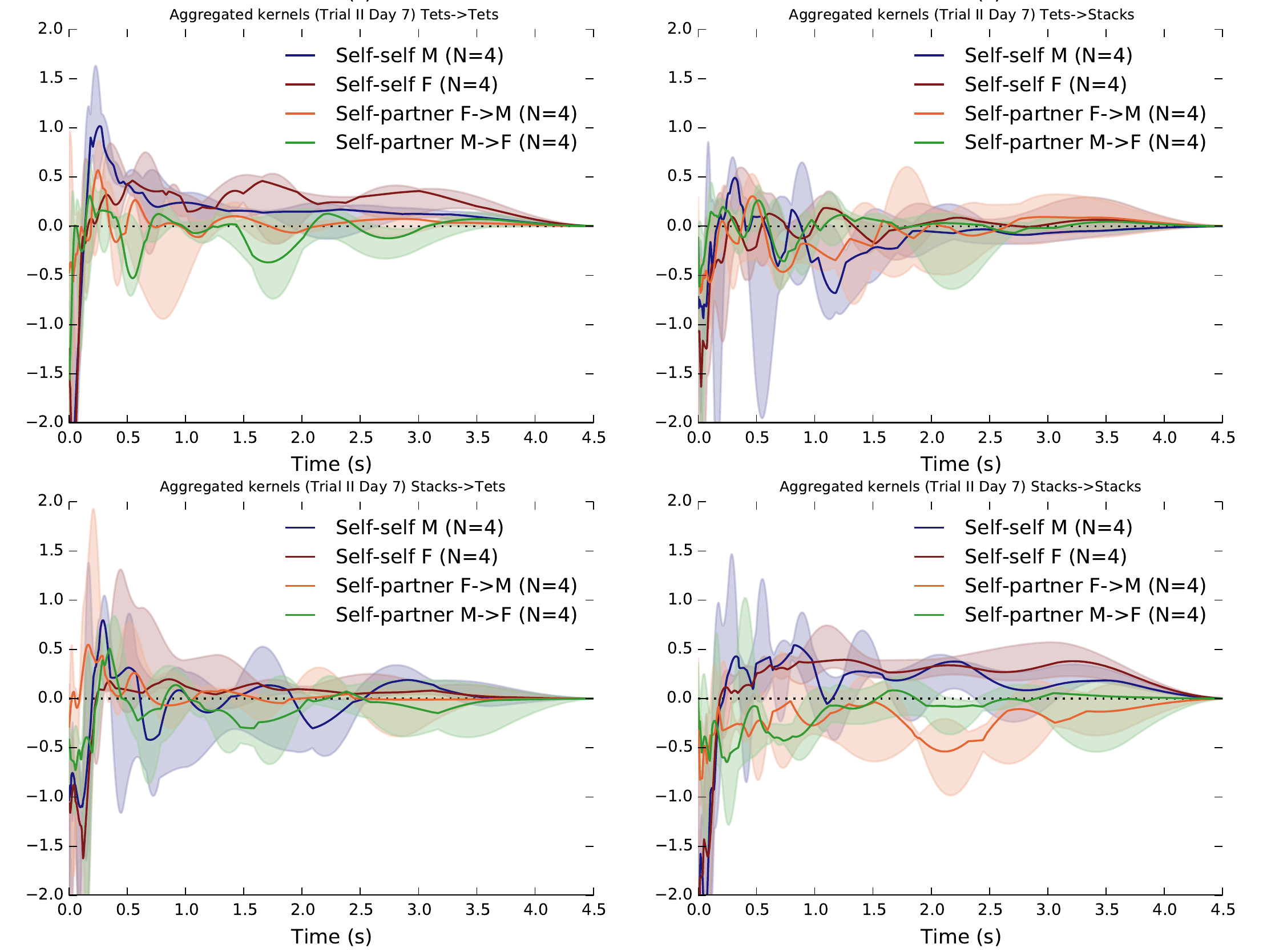}
	\label{fig:plot_kernels_ssso_gill_ind_4type_tetstack}
\end{figure*}

\begin{figure}[pt]
	\caption{Aggregate kernels specifically for the ``Cackles'' of \gill\ on day 7.}
	\centering
	\includegraphics[width=0.6\linewidth,clip,trim=340mm 87mm 110mm 254mm]{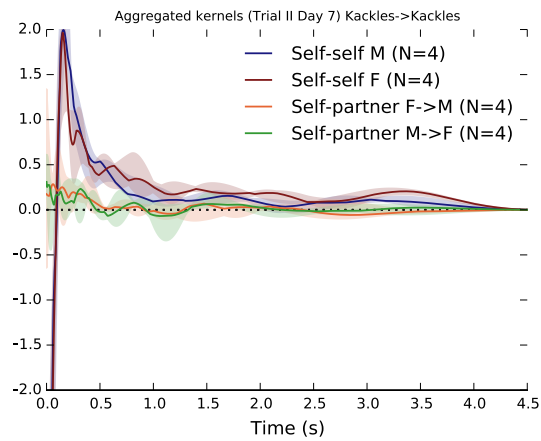}
	\label{fig:plot_kernels_ssso_gill_ind_4type_cackle}
\end{figure}

Other aspects of the kernels broken down by call type confirm the observations of \gill:
In many cases the strongest effect of a particular call type was to induce responses of the same type,
but with some influences from one call type to another.
Around day 7 we observed a tendency for Stacks or Tets from a female to induce Tet responses in the male partner (Fig. \ref{fig:plot_kernels_ssso_gill_ind_4type_tetstack}),
as was also remarked upon in \cite{Gill:2015},
and for the male Tets to have a notable self-excitatory peak.
By day 7 birds were largely in the nest-building and later nesting stages (Fig. 2 of \cite{Gill:2015}).

Note that dividing the calls into five types gives a 25-fold increase in the number of influence kernels to be recovered,
which may lead to data sparsity in some cases.
This is visible in the increased variance of the per-type kernel estimates
(Fig. \ref{fig:plot_kernels_ssso_gill_ind_4type_tetstack}, Fig. \ref{fig:plot_kernels_ssso_gill_ind_4type_cackle}).
For this reason we will only discuss per-type kernels in which we observe clear patterns,
or indicative patterns which triangulate against observations made in related work \cite{Gill:2015,Elie:2015b}.

\section*{Discussion}
\label{sec:discussion}

When animals interact in groups, multiple influences converge on each individual in parallel,
and the effects of these influences depend on social context.
In groups of zebra finches we found temporal interaction patterns that were consistent:
they persisted over time according to sender/receiver identity,
and they had characteristic structure depending on the nature of the social bond (self-self, self-partner, self other; male-female, female-male)
and on contextual status such as breeding stage.

We characterised these communication networks by fitting a simple general-purpose model which takes account of the parallel known influences converging on an individual.
The model is flexible enough to represent a wide range of pairwise influence patterns (kernels),
including kernels which show patterns of suppression \textit{and} excitation together, depending on relative timing.

When reanalysing call data previously presented by \gill \cite{Gill:2015}
in which cross-correlation had been used to identify interactions,
we were able to confirm many of the observations,
but demonstrated in more detail the temporal structure of interaction patterns, including self-self interactions.
In some social contexts (particularly nest-building),
within-pair influences dominate and calling patterns become much less strongly influenced by extra-pair group members.
The structure of the communication network changes qualitatively through the different stages of zebra finch breeding activity.

We also found qualitative differences in communication influence patterns between our female-only group and the mixed-sex group of \cite{Gill:2015}.
Note that the groups differed in various ways (presence/absence of mates, ability to interact physically, group size, backpack microphones)
and so we do not here attribute the differences between the two studies to specific contextual parameters;
that remains for future study.
However, the per-sex and per-type plots strongly suggest that sex differences in timing patterns,
and the presence of specific self-partner interactions, are the dominant factors in the differences seen between the two studies.

\subsection*{Zebra Finch Call Types and their Use in Vocal Interactions}

Our reanalysis adds extra detail to the use of the different call types recorded and studied in \cite{Gill:2015}.
An example is the specific self-excitation pattern observed for Cackle calls (Fig. \ref{fig:plot_kernels_ssso_gill_ind_4type_cackle}).
This is not discussed in \cite{Gill:2015} because the study focused only on inter-individual timings.
The specific pattern we observed corresponds with behavioural observations in the literature:
``[Cackle] calls are emitted in sequence either by one single partner (especially by the male when leading the nest search; Zann 1996a, b) or by both birds that are then performing soft duets using these calls in combination with Tet calls (Elie et al. 2010)'' \cite{Elie:2015b}.
As well as the self-excitation pattern, we found a short-term self-partner excitation effect, corresponding to the duetting mentioned.

Elie et al. \cite{Elie:2015b} take issue with the categorisation of Tet/Stack calls used by \cite{TerMaat:2014,Gill:2015}.
They propose that the ``Stack'' of \gill\ is not the ``Stack'' which Zann observed in wild zebra finches \cite{Zann:1996},
but rather that it is a variant of the ``Tet''.
They argue that the ``Tet'' and ``Stack'' are used in very similar behavioural circumstances,
and so should all be considered under the general category of ``Tet''.
On the other hand, their own acoustic analysis finds them to be similar but distinct clusters,
and so they suggest they could be referred to as ``Tet-M'' and ``Tet-S''
to avoid confusion with the ``Stack'' of Zann.

In this light, our analysis may help to illuminate whether the two categories annotated in \cite{Gill:2015} show different interaction patterns.
The question is whether the two call types are behaviourally equivalent.
If this were the case, we would certainly expect the Tet$\to$Tet and Stack$\to$Stack kernels to have similar characteristics.
We might also expect the cross-type influence kernels (Tet$\to$Stack and Stack$\to$Tet) to be broadly similar.
Note that we would not necessarily expect the cross-type kernels to look the same as the within-type kernels:
for example the cross-type kernels might show smaller influence in the hypothetical case that Tets and Stacks are behaviourally equivalent but emitted in different states of arousal, and therefore unlikely to happen in close temporal proximity.

Contrary to this hypothesis of equivalence, we found indications of differing influence kernels both within and between Tets and Stacks.
This was particularly seen on day 7 (Fig. \ref{fig:plot_kernels_ssso_gill_ind_4type_tetstack}) in which we saw that both Tets and Stacks from a female showed a tendency to inspire a Tet response from a partner within around 0.25 seconds, but this was not seen for Stack responses.
Stacks and Tets also showed differing self-excitation patterns, indicating that the short-term sequencing of bursts had different character.
Our analysis does not disprove the claim of Elie et al. \cite{Elie:2015b} that Tets and Stacks lie on a continuum and are used in similar situations.
However it indicates that as well as having observable acoustic differences, Tets and Stacks are used differently within communication interactions on the time-scale of seconds.

Elie et al. \cite{Elie:2015b} describe the Tet call thus:
``The Tet call is the most frequent vocalization as it appears to be produced in an almost automatic and continuous fashion when zebra finches move around on perches or on the ground. These ``background'' Tet calls form an almost continuous hum and do not appear to produce a particular response in the nearby birds.''
Contrary to this, we find that Tets do have an effect of inducing Tets from a partner on a specific timescale.
In this we concur with \gill \cite{Gill:2015}.

It is worth noting that in our presentation we have not focussed on the resting ``base rate'' of calling,
which in our model is the component that causes birds to call in absence of any stimulus.
The base rates here took values of approximately $\pm 0.15$ per individual;
the peak influence spikes were on a similar or larger scale and thus had a non-trivial effect compared against the base rate.
A particular appeal of our modelling approach here is that it can identify components of influence even in the presence of a base calling rate.

\subsection*{Reflections on GLM Point Process Methodology}

The GLM point process method we have used is relatively generic---it can be applied to neurons as much as to calling animals---%
and as we have shown, it is flexible enough to capture a variety of phenomena which are pertinent to the understanding of animal calling interactions.
It can capture specific timescales and strengths of influence, both positive and negative and mixed,
between individuals, including asymmetric influences (A$\to$B can be different from B$\to$A),
and provides a useful representation separating specific influences out from the calling base rate.
It can reproduce bursty/sequential calling phenomena in individuals or groups.
The method has a number of advantages over cross-correlation analysis.
Directed causation is directly modelled rather than implicit.
(This should not be interpreted as claiming that the method uncovers the full set of factors having causal influence on an individual:
the model abstracts over physiological detail, and characterises the relative strengths of the causal factors proposed by the analyst.)
Multiple convergent influences are simultaneously modelled as well as a default base-rate.
Thanks to this, spurious links due to common-cause effects are less likely to occur.

The method is not specialised for strict sequencing:
for example, if birds always emitted exactly three cackles in a sequence,
this could not be modelled.
In fact this limitation is in common with the standard Markov model.
Future modelling advances may add useful generalisations, for example the incorporation of hidden state variables.
Strict sequencing can be described in a hidden Markov model or a semi-Markov model,
but those are in general suitable only for independent individuals and not a good fit for situations with multiple influences within a group.

The point process model we have described here is closely related to a set of self-stimulating statistical models called ``Hawkes processes''
\cite{Hall:2014,Moore:2015b}.
For example, the method of \cite{Hall:2014} has an appealing property of ``streaming'' performance,
meaning that the network characteristics can adapt continuously through time as the network evolves.
However their model has important limitations which the GLM model does not.
It does not incorporate the nonlinearity which allows for flexibility and ensures that the model remains meaningful in the presence of negative influences (which otherwise could yield meaningless negative calling rates).
More importantly, under their model every link in the network must have the \textit{same} kernel shape and only the magnitudes can vary.
We have demonstrated clearly in this work that zebra finch calling networks require, at minimum, different kernels for self-self, self-partner and self-other interactions,
which have dramatically different shapes.

Our method models each calling individual as an inhomogeneous Poisson process, where the changes in calling rate are due to external influences.
An alternative approach is to model each individual (slightly more simply) as an inhomogenous Poisson process (Fig. \ref{fig:spikingprocesses}b),
and then look for correlations among their inferred underlying calling rates.
An advantage of that approach would be to accommodate smooth modulations in the base-rate of calling;
however this comes at the significant cost of probing the modelled rates only indirectly for evidence of causal influence (much as in cross-correlation analysis),
rather than directly fitting a causal model to the observed data.

Looking slightly more broadly, there are some existing methods in the animal behaviour literature that have rough analogies to our approach, but using different types of behavioural data.
Psorakis \cite{Psorakis:2015} use spatio-temporal proximity of animals as indirect indicators of affiliation, which are then used to infer a social network graph.
Note that the method there can only infer undirected (symmetric) connections between pairs of individuals, not directed connections as in the case of calling patterns here.
Another rough analogy is with \cite{Nagy:2010} who infer pigeon hierarchy from delays in flock movement responses.
In that case the observed data are continuous movement data and temporal cross-correlations in movements are the clues used, instead of discrete events, to infer networks.

The data size requirements of the GLM point process model are reasonable for our purposes,
as indicated by our ability to recover stable repeatable influence kernels with 15, 60, or 180 minutes of data.
(See also the SI for a simulation test on data size requirements).
The approach requires the same amount of data as does cross-correlation.
We note that dividing the calls down by call type as well as by individual
can lead quite quickly to data sparsity issues.
This is seen in the slightly rough nature of the median kernels and higher variance in Fig. \ref{fig:plot_kernels_ssso_gill_ind_4type_tetstack}
versus Fig. \ref{fig:plot_kernels_ssso_gill_ind}.
This is of course true for other analysis methods as well.
The GLM analysis has been applied to data sets having hundreds of neurons,
and so it has the ability to scale to larger groups than we have studied \cite{Pillow:2008}.

The computation required to fit these models is larger than to run a cross-correlation test.
In our largest data fit, analysing one of the almost four-hour sessions from \cite{Gill:2015} and breaking each of the eight individuals' calls down into five types%
---yielding a 40-by-40 fully-connected network of influences to infer---%
this took seven hours on an ordinary laptop.
Analysis can be made faster if some connections can be ruled out \textit{a priori},
such as the influence from whines to distance calls,
which we know from behavioural observations, previous work \cite{Gill:2015,Elie:2015} and the present study
do not show any notable influence.

The paradigm that we have applied in this study is relatively abstract and generic.
This has two implications.
First, it means that the fitted models can and should be compared against behavioural observations and against any more customised behavioural and/or physiological models for the species being studied, to explore the convergence of these different sources of evidence.
Second, it means that this approach is not limited to songbirds, nor to communal species, and may find application in other taxa
such as mammals or territorial songbirds.

The GLM model is generative, which allows for interesting experimental designs that can be considered in future,
such as generating large numbers of novel group call sequences as stimuli,
synthesising background ``crowd'' sounds,
or creating group interactions in which live individuals interact in real time with automatic conversational participants.

\section*{Acknowledgments}

We thank
Richard E. Turner
and Rob Lachlan
for useful discussions that influenced the work presented here.
Thanks also to Maeve McMahon for lots of assistance with the zebra finch recording study.
We thank Jonathan Pillow for making the GLM code available and for implementation advice.

\section*{Funding}
This work was supported by
EPSRC Early Career research fellowship EP/L020505/1.

\section*{Data Accessibility}

Our call timing data is publicly available via Figshare:

\begin{itemize}
\item
\zfff: https://dx.doi.org/10.6084/m9.figshare.1613791
\item
\gill\ data subset: https://figshare.com/s/73cbdf96ab156a0f0a69
\end{itemize}

Our source code to perform analyses and generate figures is available online via:
\url{https://bitbucket.org/danstowell/callnets-glm}

\section*{Competing Interests}

We have no competing interests.

\section*{Authors' Contributions}
DS conceived and designed the study, carried out the lab work (\zfff\ dataset), performed data analysis and drafted the manuscript;
LG provided data, participated in data analysis and helped draft the manuscript;
DC participated in the design of the study, provided zebra finch resources, participated in data analysis and helped draft the manuscript.
All authors gave final approval for publication.

\bibliographystyle{vancouver}
\bibliography{../../refs}

\clearpage
\appendix

\section*{Supporting Information}

\subsection*{GLM point process model fitting}

Our use of the GLM point process model broadly follows that of \pillow \cite{Pillow:2008}.
Here we describe specific configuration choices and adaptations we made to the model.

\subsubsection*{Choice of nonlinearity}

In the GLM point process model, the instantaneous firing rate $\lambda$ is given by a \textit{linear-nonlinear} link function, meaning that influences are linearly summed and then passed through a nonlinearity:
\begin{equation}
	\label{eqn:lnp}
	\lambda(t) = \sigma\left(b + \sum_i{K_i \ast y_i(t)}\right)
\end{equation}
where we have used $b$ to represent the base rate component, $K_i$ the influence kernel from individual $i$, $\ast$ the convolution operation, and $y_i(t)$ the sequence of events emitted by individual $i$ represented as a spike train.

The function $\sigma$ is a nonlinear mapping which can be freely chosen within certain constraints;
these include that the function must be monotonic and non-negative
\cite{Paninski:2004}.
If $\sigma$ is the exponential function, then this has the effect of transforming \eqref{eqn:lnp} so that the influences combine in multiplicative fashion rather than additive.
If $\sigma$ is the rectifier function $\max(0, \cdot)$, this preserves linear additivity except that negative values are clipped away.
However, the rectifier function can be difficult to perform numerical optimisation upon, because a large part of its domain gives zero gradient,
and often a smooth approximation is used in its place, which we describe shortly.

\begin{figure}[tp]
	\caption{Nonlinearity functions.}
	\centering
	\includegraphics[width=0.8\linewidth,clip,trim=0mm 0mm 0mm 0mm]{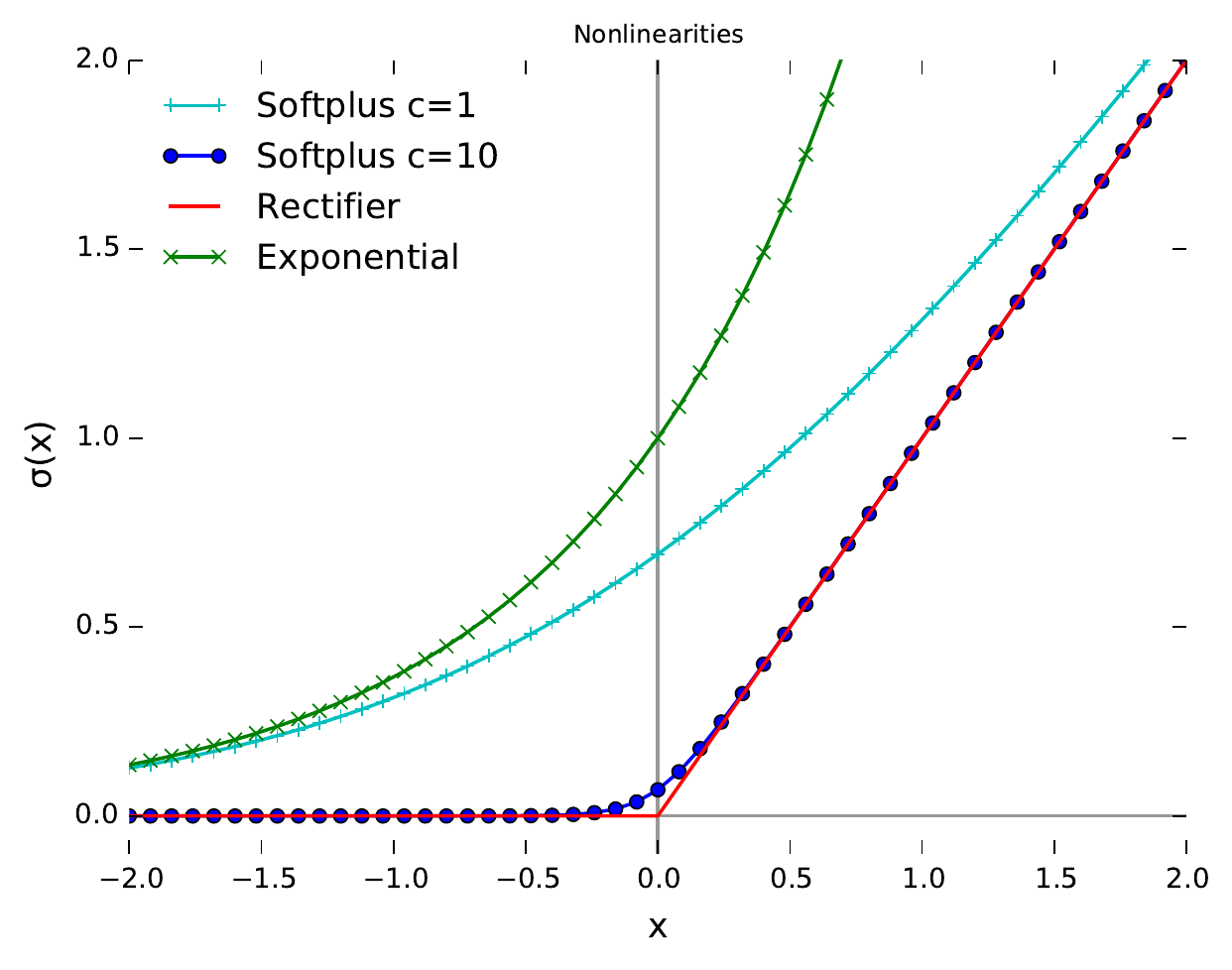}
	\label{fig:plot_nonlins}
\end{figure}

Within this simple model, in which animal behaviour is embedded in rather abstract form, it is not immediately clear which nonlinearity should be chosen.
(Note that the method is rather robust to misspecification, meaning that under mild conditions it yields consistent kernel estimates even if the wrong link function is used \cite[Section 5]{Paninski:2004}.)
Hence we chose to use these simple nonlinearities and use model comparison to choose between additive and multiplicative models, as was done in \cite{Pillow:2008}.
In early tests we found that the strict rectifier function often failed in optimisation, and so in its place we used the smooth \textit{softplus} function
\begin{equation}
	\lambda(x) = \log(1 + \exp(cx))/c
\end{equation}
where we introduced the constant scaling factor $c$ to bring the function closer to the rectifier nonlinearity.
The softplus function is not scale-invariant and so its effect (and hence the desirable choice of $c$) depends on the dynamic range of the data.
We used $c=10$, chosen to be large enough to bring the function close to a true rectifier effect without leading to numerical optimisation problems (Figure \ref{fig:plot_nonlins}).

When given an input of zero, our softplus nonlinearity yields a rate of around 0.07 (one call expected every 15 seconds).
Negative inputs suppress the rate even further: in theory the rate never quite reaches zero, while with the machine precision used here, a zero rate of emission is produced from an input of around $-4$ or less.
Positive inputs become increasingly close to the identity mapping; any input larger than around 0.3 is within 1 per cent deviation from identity.

Typical values of the base-rate parameter $b$ in fitted models were approximately around $\pm 0.15$.
The inter-individual variation in $b$ was smaller than the general scale of the influence kernels.

\subsubsection*{Kernel basis functions}

\begin{figure}[tp]
	\caption{The raised-cosine basis functions used to compose each kernel.}
	\centering
	\includegraphics[width=0.8\linewidth,clip,trim=0mm 0mm 0mm 0mm]{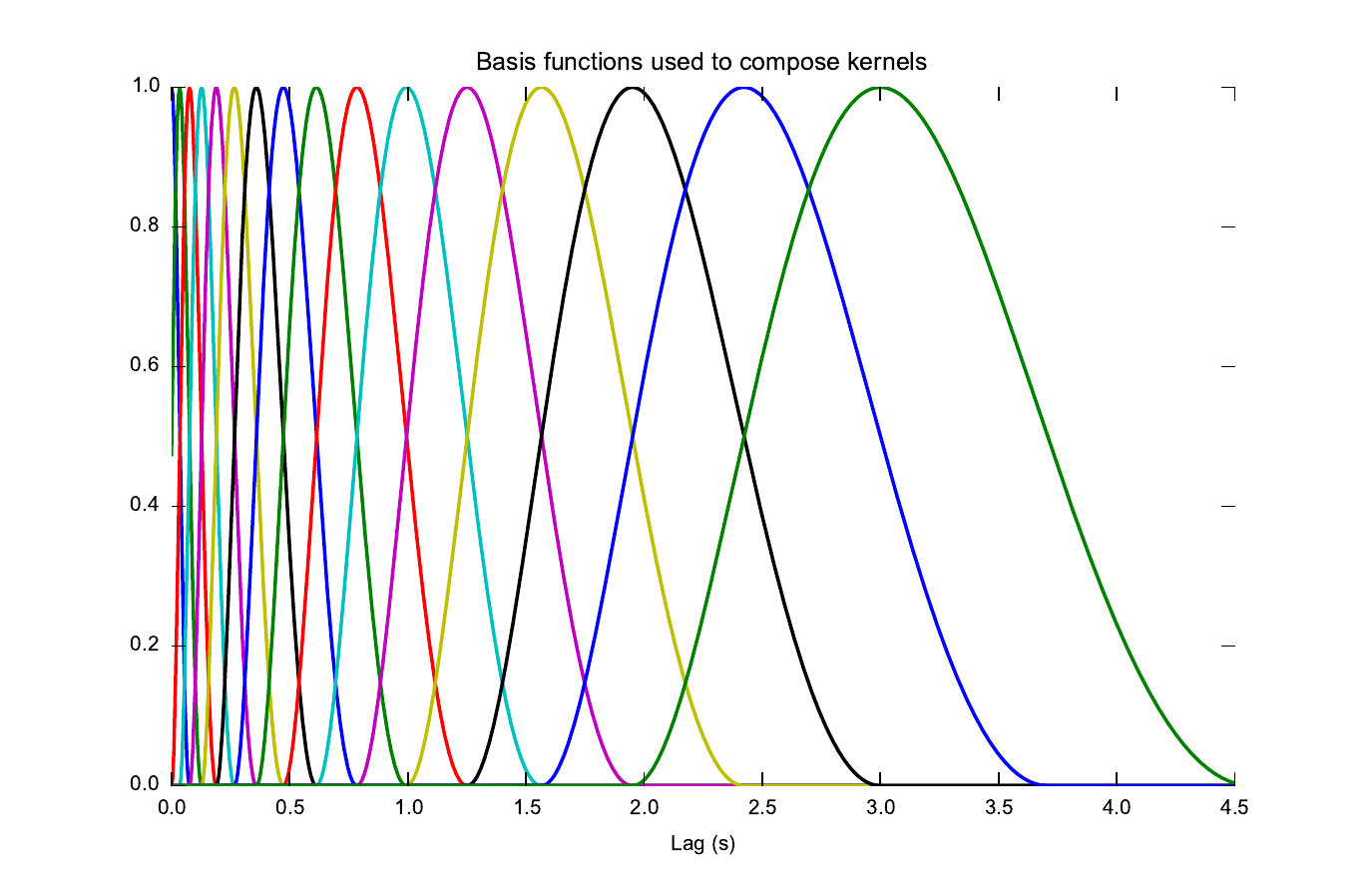}
	\label{fig:plot_basisfuncs_r}
\end{figure}

The model works by parametrically combining a set of basis functions to create each kernel.
This keeps the dimensionality of the model tractable while allowing for great flexibility in the range of kernels possible.
We used a set of raised cosine functions (Figure \ref{fig:plot_basisfuncs_r}),
distributed in a warped arrangement to offer more detail at short time lags.
In initial tests we first used 8 basis functions.
We varied the warping factor to inspect whether the choice of bases had a strong influence on the resulting kernel fits.
We found that with 8 basis functions the placement of the bases had an observable effect on the undulations of the median fitted kernels,
but if we increased to 16 basis functions the recovered kernels were much more stable to the exact choice of basis.
We therefore used 16 basis functions for the main analyses.
The same set of basis functions was used for all network connections, including self-self and self-other.

\subsubsection*{Regularisation}

\begin{figure}[tp]
	\caption{Ten randomly-sampled examples of kernels from the prior distribution of the GLM model (i.e. without any exposure to data), to illustrate the range of kernels that can be fitted and also some qualities of the prior.
Note the symmetry of the prior distribution around zero, and the relative smoothness of kernels at long lag times---because the basis functions are distributed exponentially with respect to lag time.}
	\centering
	\includegraphics[width=0.8\linewidth,clip,trim=0mm 0mm 0mm 0mm]{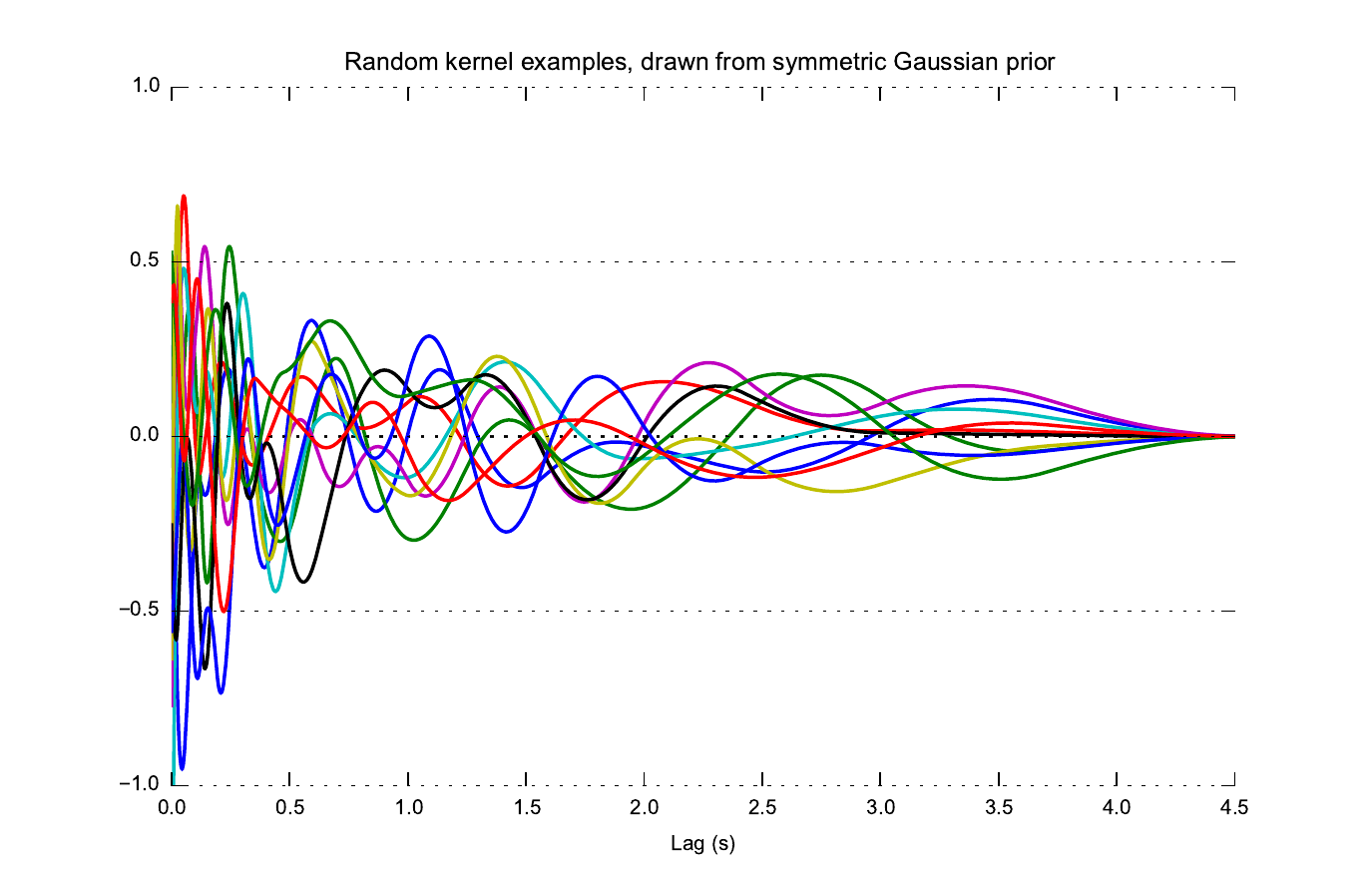}
	\label{fig:plot_basisfuncs_s}
\end{figure}

The model fitting method finds a single point estimate of the best fitting parameters.
This is done by maximum likelihood in \cite{Pillow:2008},
but it can also be done by maximum \textit{a posteriori} (MAP; also referred to as a type of penalised maximum likelihood) within the same framework
\cite{Paninski:2004}.
It is widely known that maximum likelihood methods are vulnerable to overfitting in the presence of limited data.
We therefore used MAP regularisation in the GLM model to help prevent overfitting at small data sizes and to stabilise the fitted parameter estimates.
We used standard $L_2$ (Euclidean) regularisation when fitting the model, which is equivalent to a zero-centred Gaussian prior on the coefficients (Figure \ref{fig:plot_basisfuncs_s}).
We did not regularise the parameter $b$---the fixed base-rate for each individual to emit an event in absence of stimulation---allowing it to take any value with equal prior probability.

\subsection*{Synthetic A$\to$B$\to$C example and data size tests}
\label{sec:abc}

The A$\to$B$\to$C synthetic example given in Figure \ref{fig:abcexample} was generated from a simplified model that is not an exact match to the GLM or MRP generative models,
in order to provide a clear example of indirect causation and to examine how the various analyses behave.
First, `A' events were generated from a homogeneous Poisson process with fixed rate of 0.1 Hz.
Then, for the two links in the causal chain, A$\to$B and B$\to$C,
each possible source event generated a target event with a probability of 0.9,
and if it did so, with a time gap drawn from a log-Gaussian distribution with log-mean 0.75 and log-standard deviation 0.5 (using the natural logarithm).
The parameter values here were manually chosen to be on a similar timescale as zebra finch calls,
but not intended to be a likely zebra finch scenario, rather to demonstrate the common-cause phenomenon.
The generative model just described does not quite match the GLM point process model since no more than one event can be generated as a result of any other event, and there is no nonlinearity in the influence on rates.
Nor does it match a pure Markov model: although each A$\to$B$\to$C sequence considered on its own is Markovian, the process can generate overlapping sequences meaning the overall sequence has no upper bound on its history dependence.
This was therefore a useful test of how the different analysis procedures respond to data coming from a simple causal chain.
Using this procedure, a timeline of length 96,000 seconds was generated.

For cross-correlation, we used a custom Python script to implement cross-correlation with a maximum lag of 2.5 seconds.
Spike data were smoothed using a Hann window of duration 200 ms.

\paragraph{Data size tests}

In order to get a practical impression of the data size requirements of the GLM method compared against cross-correlation,
we applied both methods to the synthetic A$\to$B$\to$C model
while varying the number of data points fed into the analysis.
This way we could guarantee that the source of data was a stationary system with unchanging characteristics,
and inspect convergence as data size increased.

The analysis models are not designed to recover the same information, so they are not suitable for numerical comparison (e.g. mean squared error).
Under the A$\to$B$\to$C model there are six inter-individual interactions to analyse,
or three if undirected pairs are considered.
We visualised the fitted parameters by allocating each of the three pairings to a colour (red/green/blue) and superimposing the three cross-correlation curves.
The GLM model recovers six inter-individual kernels so we concatenated each kernel with its kernel in the opposite direction, to produce three shapes that could be colour-mapped in the same way.

\begin{figure}[tp]
	\caption{Visualisations of the cross-correlation curves and the inter-individual influence kernels recovered from data generated by the ABC model, as a function of the volume of data fed in to the algorithms. Plots show cross-correlation (upper), GLM (middle) and GLM without regularisation (lower). In each case the kernels/cross-correlations are superimposed as channels of an RGB image, and then normalised for dynamic range. The lower plot appears bolder because the unregularised GLM tended to dramatically large parameter values when overfitting to small datasets.}
	\centering
	\includegraphics[width=0.8\linewidth,clip,trim=0mm 0mm 0mm 110mm]{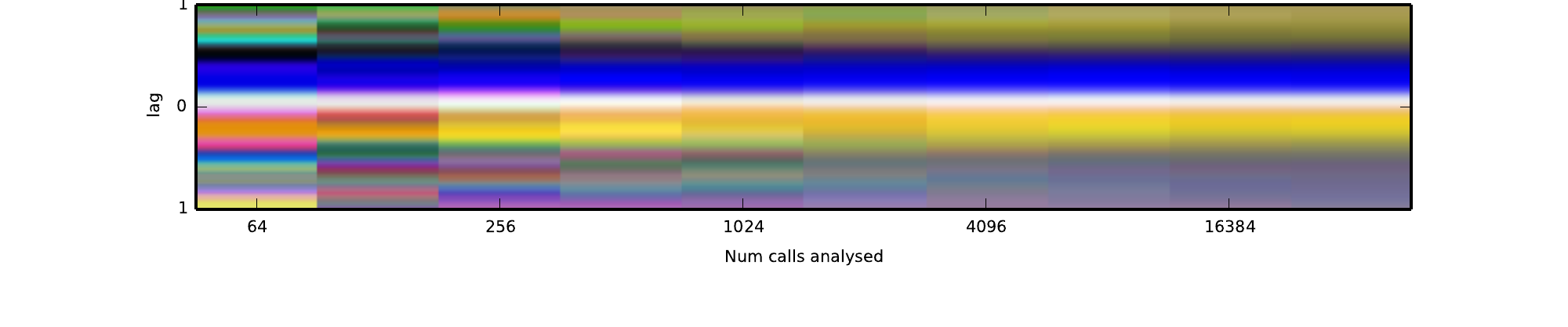}
	\includegraphics[width=0.8\linewidth,clip,trim=0mm 0mm 0mm 110mm]{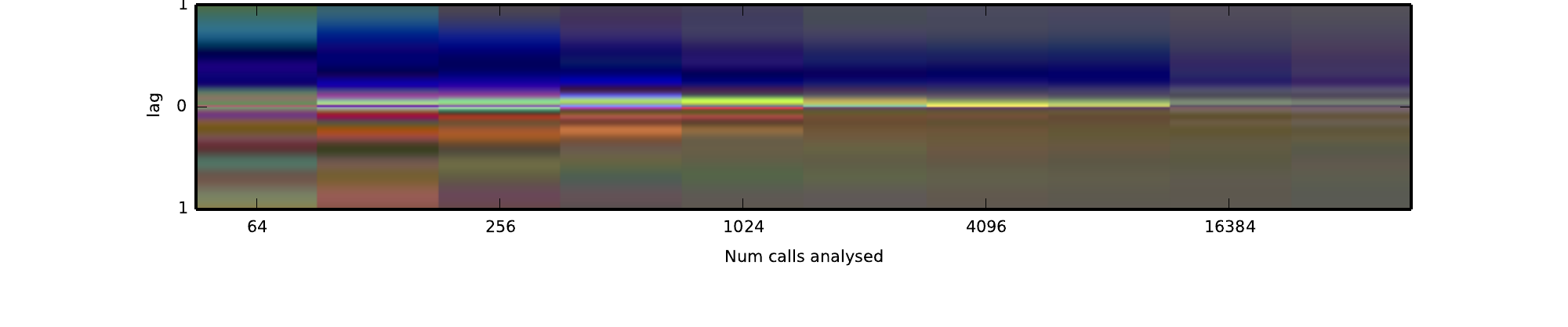}
	\includegraphics[width=0.8\linewidth,clip,trim=0mm 0mm 0mm 110mm]{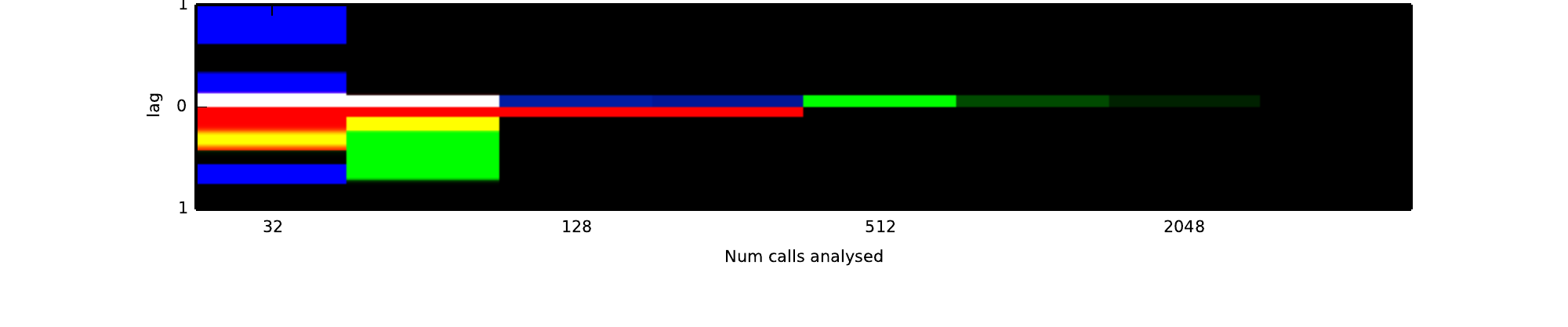}
	\label{fig:datasizeconvergence}
\end{figure}

We visualised the parameters recovered from the cross-correlation and GLM models, and also the GLM model without regularisation (i.e. maximum likelihood rather than maximum \textit{a posteriori} fitting) (Figure \ref{fig:datasizeconvergence}).

We found that both cross-correlation and the GLM model converged at approximately 1024 data points.
As expected, the use of regularisation in the GLM model helped prevent overfitting at small data sizes, and to stabilise the fitted parameter estimates.
We therefore used this same regularisation for all fits reported in the paper.

\subsection*{Simulating from the fitted models}

Since our model is generative, we can draw newly simulated event sequences from the fitted models,
which will thus be qualitatively similar to the original material.
These sequences can be of arbitrary length and could even be generated in online (real-time) fashion.
We can also use resimulation as another approach to inspecting and validating the fitted models.

\begin{figure}[tp]
	\caption{Timeline plot of an excerpt from Session 2 of \zfff.}
	\centering
	\includegraphics[width=0.89\linewidth,clip,trim=20mm 10mm 18mm 10mm]{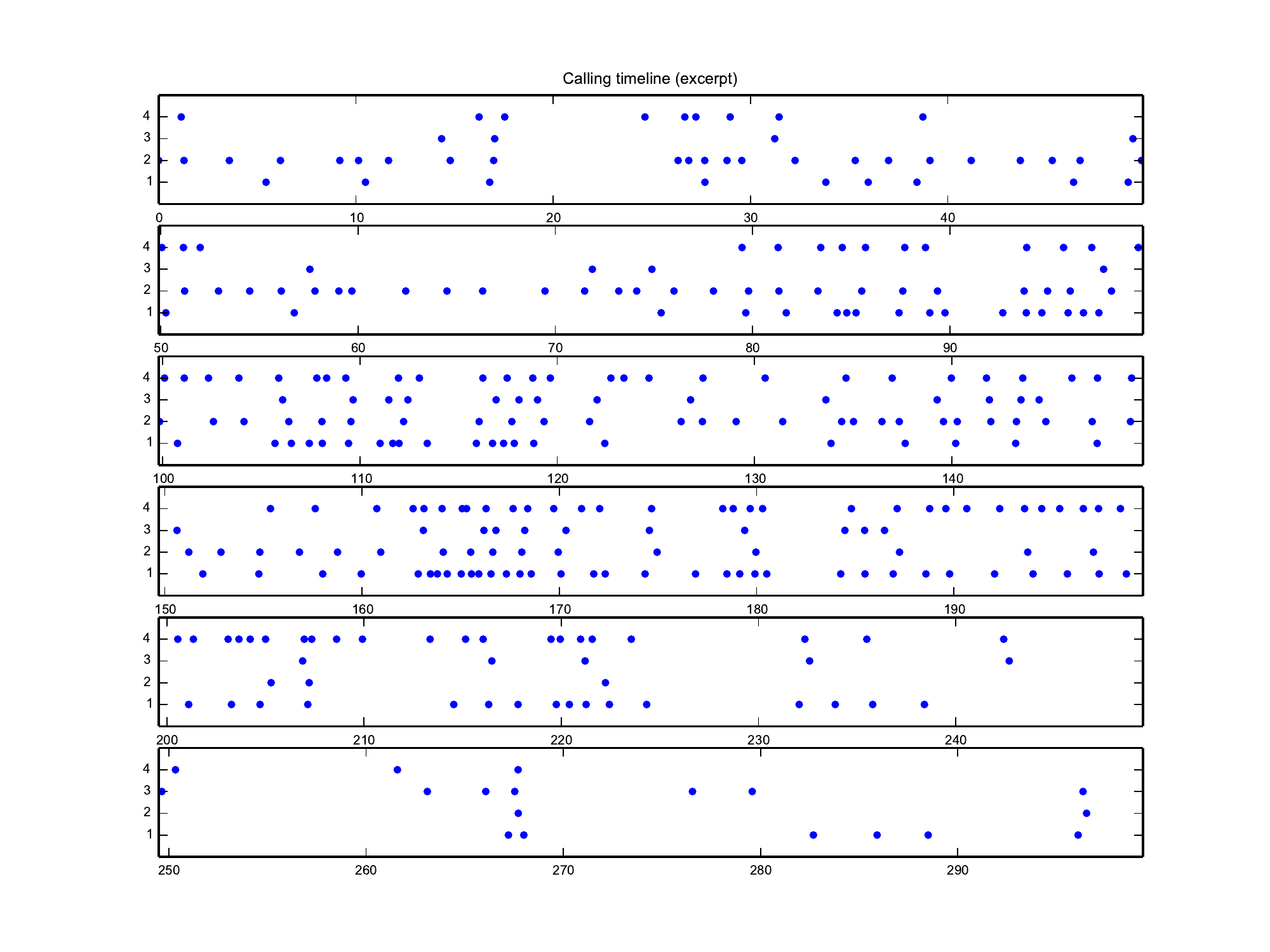}
	\label{fig:plot_timeline_session2a}
\end{figure}

\begin{figure}[tp]
	\caption{Timeline plot of artificial data resimulated from fitted model for Session 2 of \zfff.}
	\centering
	\includegraphics[width=0.89\linewidth,clip,trim=20mm 10mm 18mm 10mm]{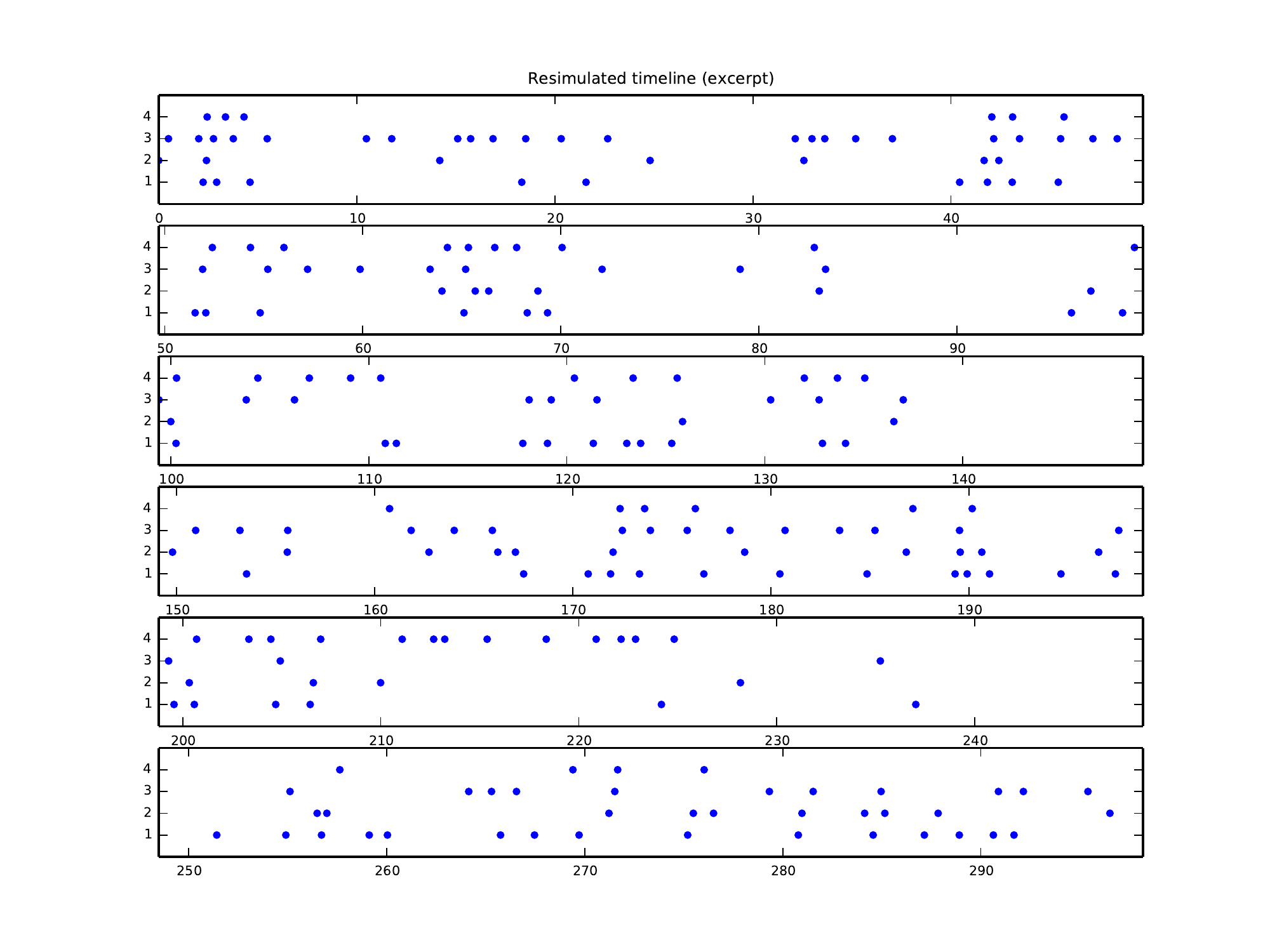}
	\label{fig:plot_timeline_session2fullsof_resimulated}
\end{figure}

\begin{figure}[tp]
	\caption{Timeline plot of artificial data resimulated from fitted model for Session 2 of \zfff, in this case with only one virtual bird and no group.}
	\centering
	\includegraphics[width=0.89\linewidth,clip,trim=20mm 10mm 18mm 10mm]{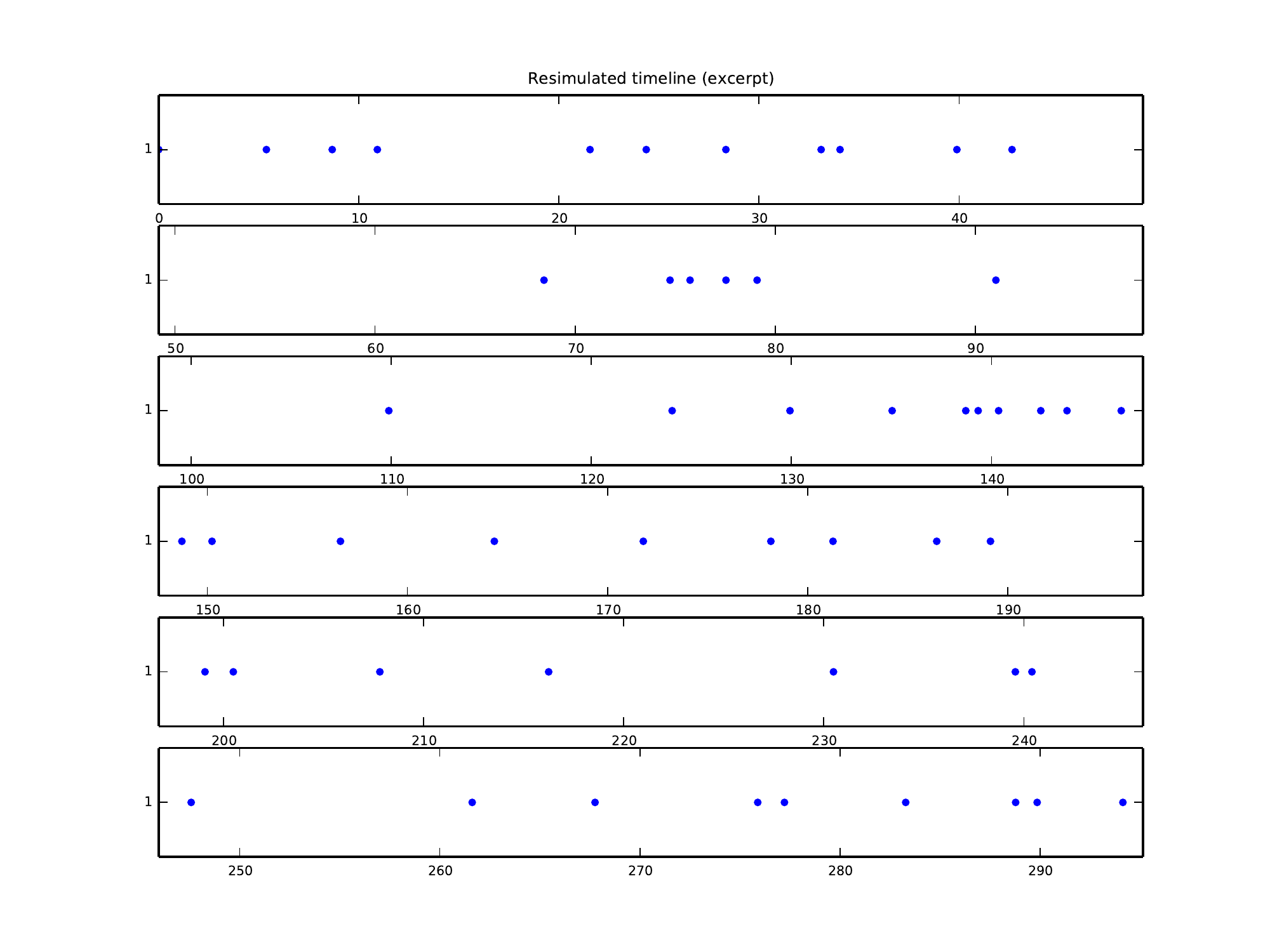}
	\label{fig:plot_timeline_session2fullsof_resimulated_asolo}
\end{figure}

As a simple illustration of this, we resimulated 600 seconds of calls using the models fitted to the \zfff data.
We resimulated the group (Figure \ref{fig:plot_timeline_session2fullsof_resimulated}),
and also we resimulated a single individual taken out of the group context,
i.e.\ with only base-rate and self-self effects (Figure \ref{fig:plot_timeline_session2fullsof_resimulated_asolo}).
Between these two different simulations, the effect of group interactions was visible in the data:
the same virtual individual produced calls at a rate of one per 14 seconds when simulated alone,
and at a rate of one every 6.5 seconds when simulated in the group.
(The real data contained a call from that individual every 4.6 seconds.)
The resimulated group data replicates qualitative aspects of the real data (Figure \ref{fig:plot_timeline_session2a})
such as the bursty communal calling, though appearing slightly less dense.


Re-running the GLM fitting procedure on the resimulated data for the group recovered kernels with the same shapes as in the source model (Figure \ref{fig:plot_kernels_ssso_resim}).

\subsection*{Further plots from datasets}

\begin{figure}[tp]
	\caption{Density plots of the calling rates of each bird in the \zfff recording study. Rates are calculated from call times using kernel density estimation with a bandwidth of 15 seconds.}
	\centering
	\includegraphics[width=0.8\linewidth,clip,trim=0mm 0mm 0mm 0mm]{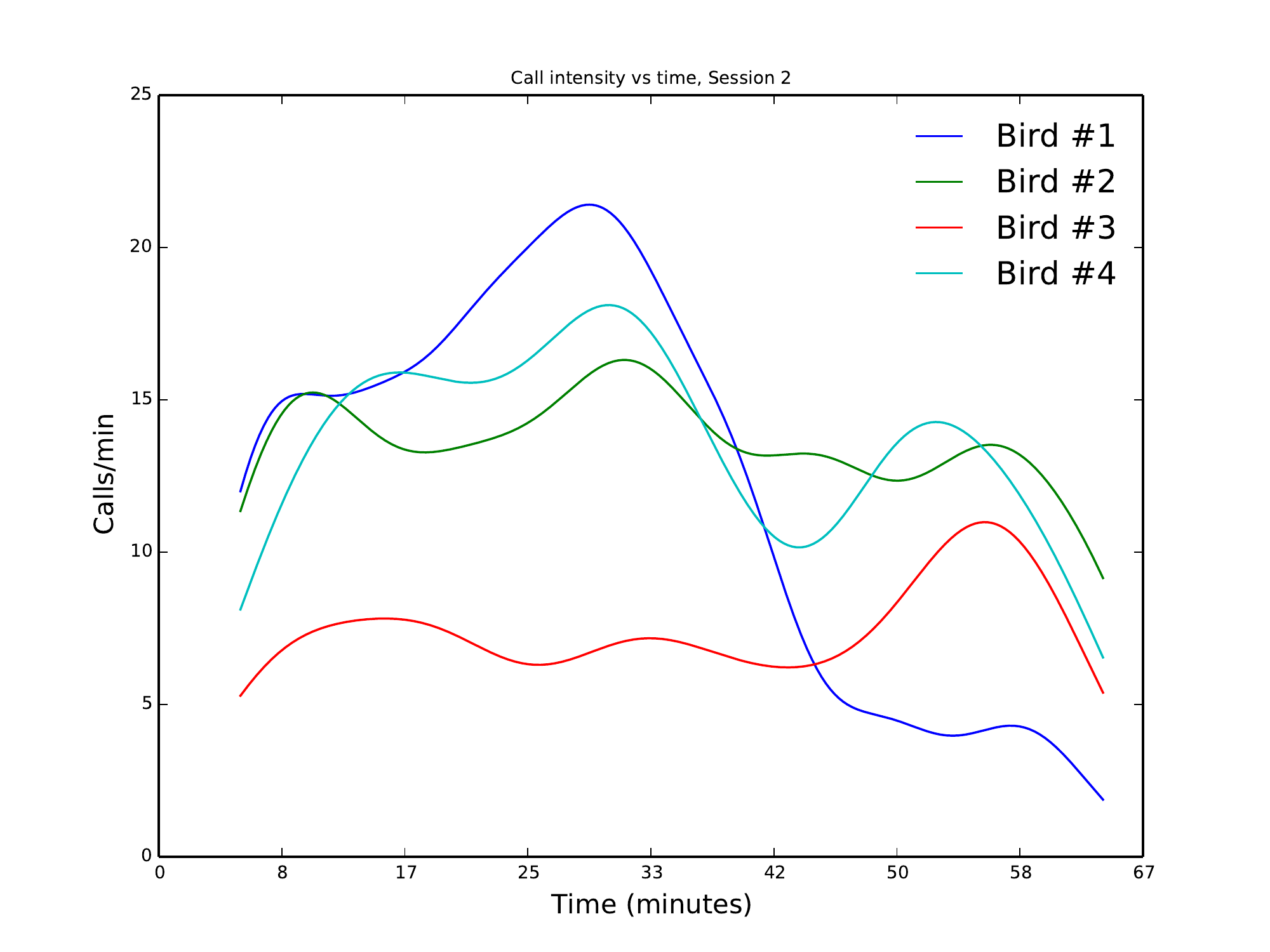}
	\includegraphics[width=0.8\linewidth,clip,trim=0mm 0mm 0mm 0mm]{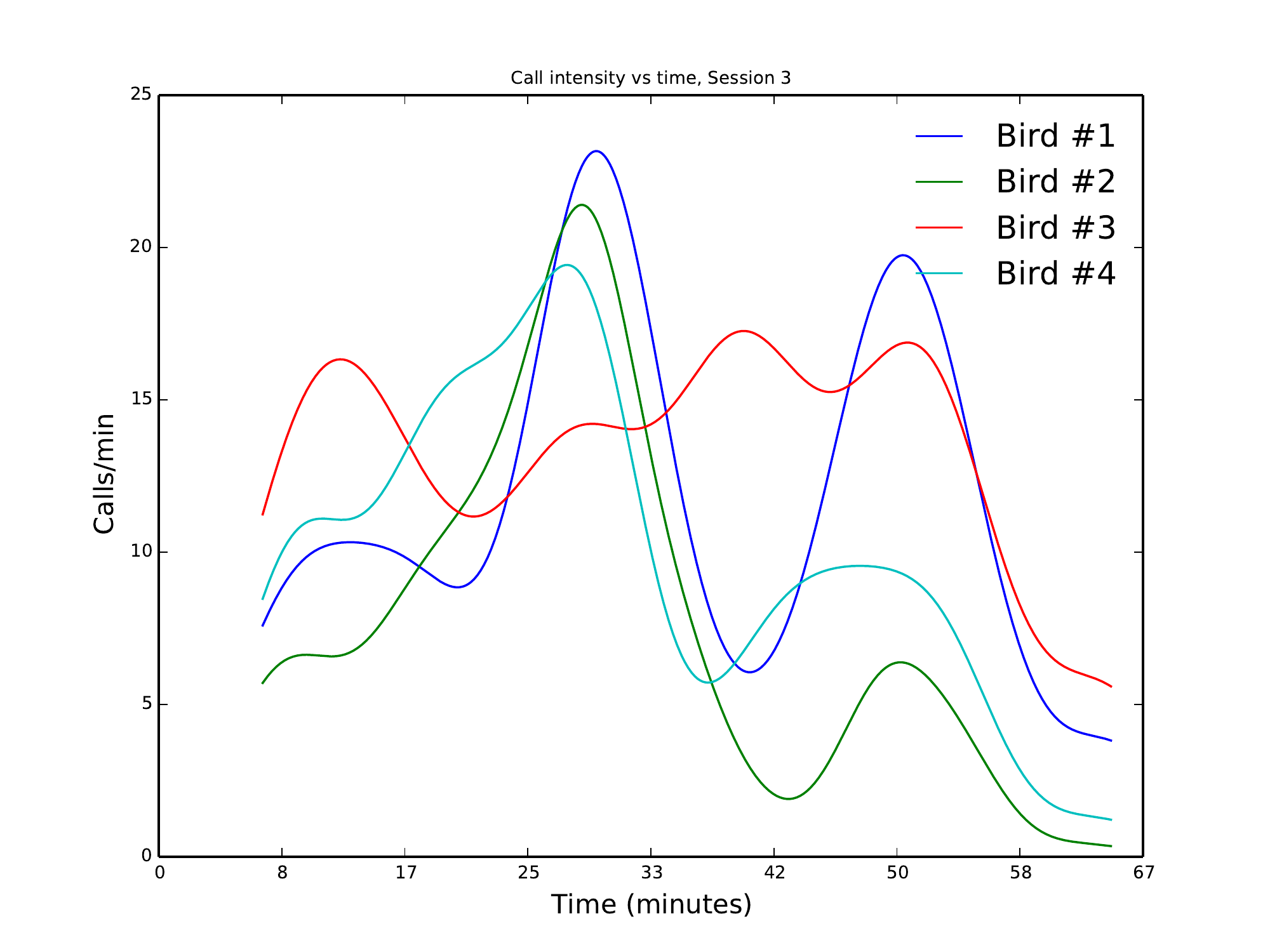}
	\label{fig:kdeplot}
\end{figure}

\begin{figure*}[pt]
	\caption{Kernels recovered from the resimulated data.}
	\centering
	\includegraphics[width=0.89\linewidth,clip,trim=0mm 0mm 0mm 0mm]{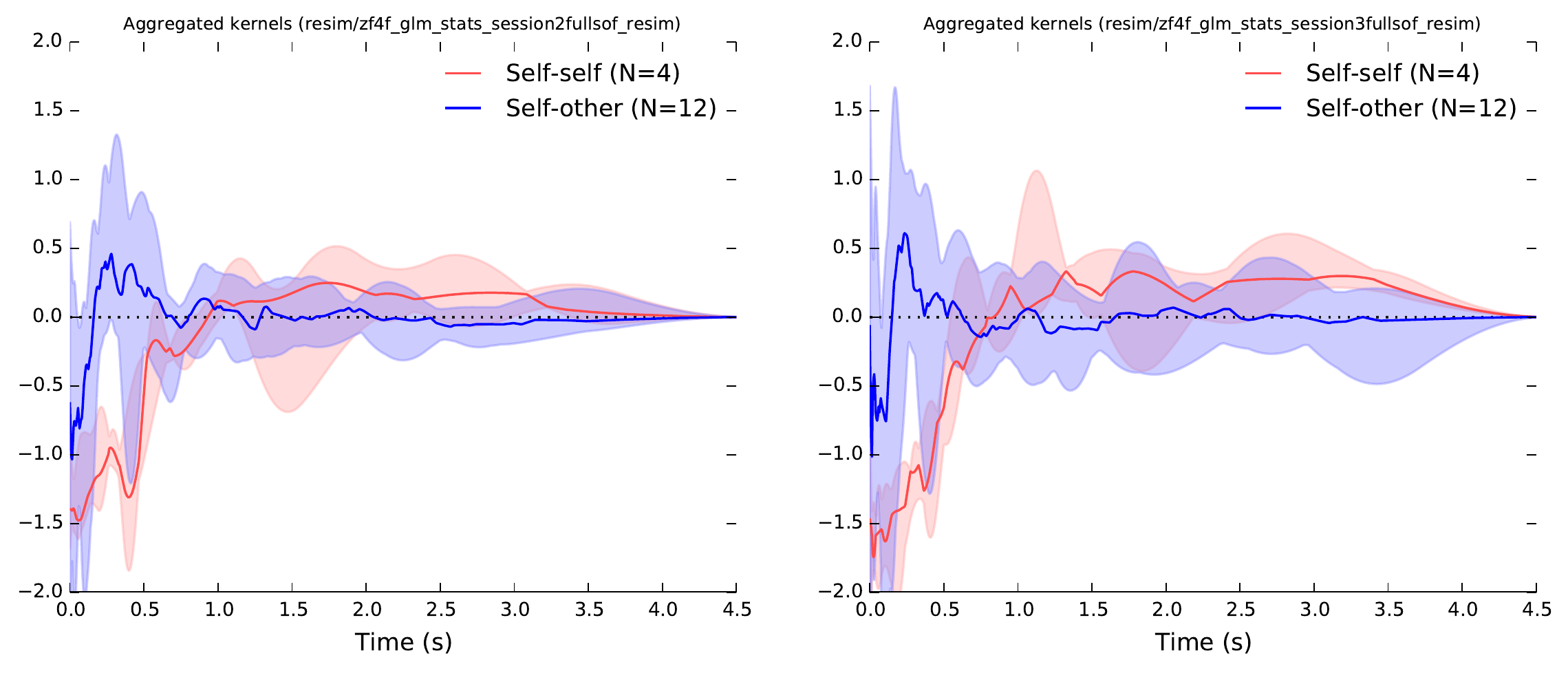}
	\label{fig:plot_kernels_ssso_resim}
\end{figure*}

\begin{figure*}[tp]
	\caption{Aggregate view of the influence kernels recovered from our two study sessions with four female zebra finches, as Figure \ref{fig:plot_kernels_ssso} but showing every one of the 16 curves fitted to each session, rather than confidence intervals.}
	\centering
	\includegraphics[width=0.89\linewidth,clip,trim=0mm 0mm 0mm 0mm]{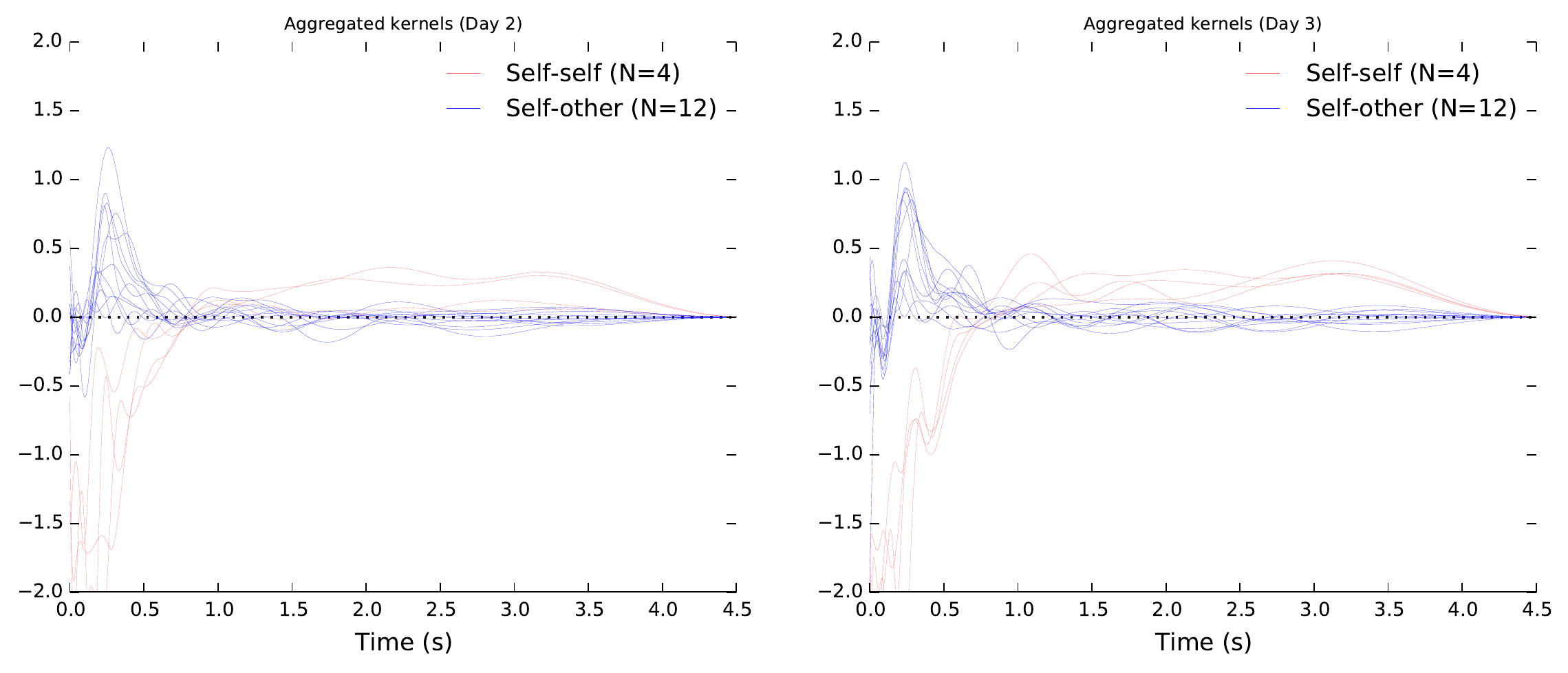}
	\label{fig:plot_kernels_ssso_allcurves}
\end{figure*}

For the \zfff dataset, overall calling rates per individual ranged from 468 to 888 calls per hour (Table \ref{tbl:callcounts}),
and rates exhibited some modulation across each hour (Figure \ref{fig:kdeplot}).
A composite of all the influence kernels fitted to the dataset is shown in Figure \ref{fig:plot_kernels_ssso_allcurves}.

\begin{table}[tp]
	\caption{Total calls emitted by each bird in the \zfff recording study.}
	\centering
	\begin{tabular}{rrr}
Individual & Calls in Day 2 session & Calls in Day 3 session \\
\hline
1 & 761 & 736 \\
2 & 888 & 468 \\
3 & 481 & 850 \\
4 & 856 & 612 \\
	\end{tabular}
	\label{tbl:callcounts}
\end{table}

\begin{figure*}[pt]
	\caption{Aggregate kernels for all possible call-type interactions, for the data of \gill on day 7.}
	\centering
	\includegraphics[width=0.999\linewidth,clip,trim=0mm 0mm 0mm 0mm]{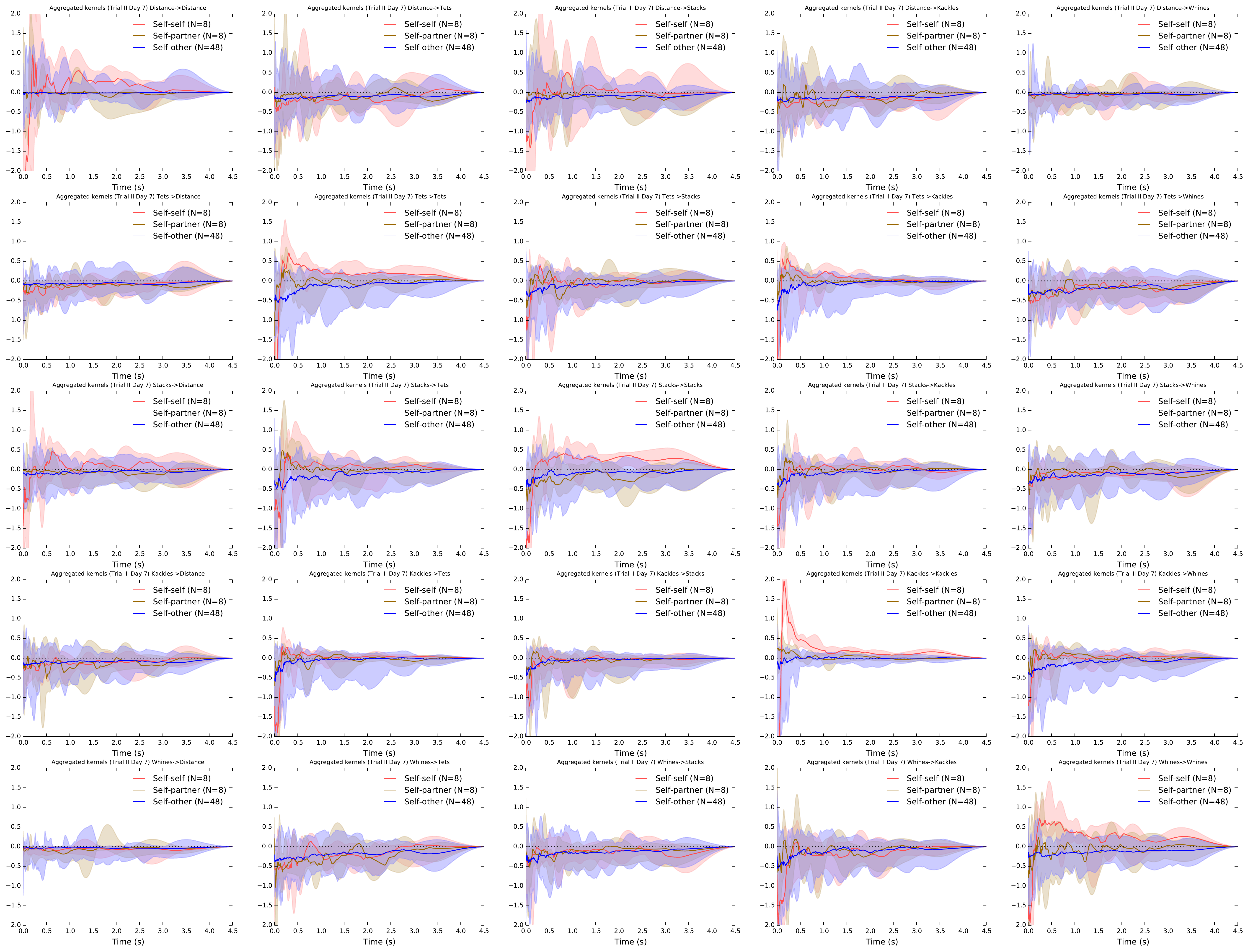}
	\label{fig:plot_kernels_ssso_gill_ind_4type}
\end{figure*}

For completeness we show the fitted kernels for all possible call-type interactions, for a specific day of interest in the \gill dataset (Figure \ref{fig:plot_kernels_ssso_gill_ind_4type}).

\clearpage

\end{document}